\documentclass[%
 reprint,
 amsmath,amssymb,
 aps,
]{revtex4-2}

\usepackage[pdftex]{graphicx}

\graphicspath{
{./Figures/},
{./Figures/Bands/},
{./Figures/DBSPs/},
{./Figures/Phonons/},
{./Figures/Prototypes/},
}

\usepackage{diagbox}
\usepackage{braket}
\usepackage{dcolumn}
\usepackage{bm}
\usepackage{booktabs}
\usepackage{threeparttable}

\begin{document}

\preprint{APS/123-QED}

\title{Prediction of quaternary hydrides based on densest ternary sphere packings}

\author{Ryotaro Koshoji}
\email{cosaji@issp.u-tokyo.ac.jp}
\affiliation{Institute for Solid State Physics, The University of Tokyo, Kashiwa 277-8581, Japan}
\author{Masahiro Fukuda}
\email{masahiro.fukuda@issp.u-tokyo.ac.jp}
\affiliation{Institute for Solid State Physics, The University of Tokyo, Kashiwa 277-8581, Japan}
\author{Taisuke Ozaki}%
 \email{t-ozaki@issp.u-tokyo.ac.jp}
\affiliation{Institute for Solid State Physics, The University of Tokyo, Kashiwa 277-8581, Japan}

\date{\today}

\begin{abstract}

We exhaustively search quaternary metal hydrides based on the (13-2-1) and (13-3-1) 
structures that are the two of the putative densest ternary sphere packings in 
cubic systems [R. Koshoji \textit{et al.}, Phys. Rev. E \textbf{104}, 024101 (2021)]. 
The 73304 candidate hydrides are generated by substituting the small spheres with hydrogen atoms, 
the medium, large, and fourth spheres with metallic atoms. 
Especially, the substitution of the small spheres with hydrogen atoms gives the unconventional 
hydrogen sublattices. We screen unstable hydrides in the candidates through geometrical optimizations, 
constant pressure molecular dynamics simulations, and phonon calculations under hydrostatic pressure of $10$ GPa, 
and identify 23 hydrides with static and dynamic stability,  
including $\mathrm{H}_{12} \mathrm{Sc} \mathrm{Y}_{2} \mathrm{La}$ 
and $\mathrm{H}_{12} \mathrm{Ti} \mathrm{Ni}_{3} \mathrm{Ba}$.
We expect that the 23 candidates of hydrides screened by the exhaustive search 
for 73304 hydrides provide a guideline to narrow down the search space for trials in 
the experimental synthesis of quaternary metal hydrides.

\end{abstract}

\maketitle

\section{INTRODUCTION}

The recent progress of crystal structure prediction methods, including the evolutional algorithms~\cite{doi:10.1063/1.2210932, doi:10.1021/ar1001318, LYAKHOV20131172, 10.2138/rmg.2010.71.13, 10.1039/9781788010122}, particle swarm-intelligence approach~\cite{PhysRevB.82.094116, WANG20122063, WANG2016406}, and \textit{ab initio} random structure searching approach~\cite{PhysRevLett.97.045504, Pickard_2011}, has contributed a great deal to the discovery of novel materials. For example, a considerable number of binary hydrides were successfully predicted, and subsequently several cubic hydrides among them, including $\mathrm{YH}_{6}$ and $\mathrm{LaH}_{10}$~\cite{PhysRevLett.119.107001, Liu6990}, were experimentally synthesized as high-temperature superconductors (SCs) under high pressure~\cite{doi:10.1002/anie.201709970, Drozdov2019, PhysRevLett.122.027001, doi:10.1063/1.5128736, https://doi.org/10.1002/adma.202006832}. Furthermore, the prediction of the metastable hydride $\mathrm{Li}_2 \mathrm{MgH}_{16}$, which was predicted to be a room-temperature SC~\cite{PhysRevLett.123.097001}, ignites more interest in seeking possible ternary and quaternary hydrides with high-temperature superconductivity, which is expected to be synthesized under mild conditions. 
However, it remains a big challenge to study ternary and quaternary systems since the size of the chemical and configuration space becomes explosively large for these cases. In fact, the diversity of ternary hydrides makes it difficult to predict high-temperature SCs in spite of great efforts to search ternary hydrides~\cite{PhysRevB.93.224513, doi:10.1021/acs.inorgchem.6b01949, PhysRevB.96.144518, C7CP05267G, doi:10.1021/jacs.7b04456, PhysRevB.98.174101, PhysRevB.99.060102, PhysRevB.99.100505, Xie_2019, PhysRevB.100.184502, 10.1038/s41524-019-0244-6, PhysRevB.102.184103, PhysRevB.102.014516, PhysRevB.101.134504, doi:10.1021/acs.jpclett.0c02299, C9CP06008A, PhysRevB.104.L020511, https://doi.org/10.1002/qua.26459, LIU2021127109, doi:10.1021/acs.chemmater.1c02371, 10.1038/s41524-021-00691-6, PhysRevLett.128.047001, https://doi.org/10.1002/wcms.1582}, while a few high-pressure experiments also found that some ternary hydrides are SCs~\cite{doi:10.1021/acs.jpcc.5b03709, PhysRevB.99.024508, 10.1038/s41586-020-2801-z}.

The hydrogen sublattices in clathrate hydrides such as $\mathrm{YH}_{6}$ and $\mathrm{LaH}_{10}$ are predominantly responsible for high-temperature Bardeen-Cooper-Schrieffer (BCS) superconductivity since hydrogen has the lightest mass, strong electron-phonon coupling, and the large electron density of states at the Fermi level, consisting of the degenerate $1s$ states of hydrogen orbitals due to the high symmetries~\cite{PhysRev.108.1175, PhysRevLett.92.187002}. Note that metallic hydrogen is predicted to be a high-temperature phonon-mediated SC~\cite{PhysRevLett.21.1748, PhysRevLett.78.118}, but an extremely high pressure of 495 GPa seems to be necessary to synthesize the metal hydrogen~\cite{doi:10.1126/science.aal1579}. Instead, so-called {\it chemical compression} from relatively heavier elements can be utilized to stabilize the hydrogen sublattices in clathrate hydrides under a relatively low pressure~\cite{PhysRevLett.92.187002}. 

\begin{figure}
\centering
\includegraphics[width=1.0\columnwidth]{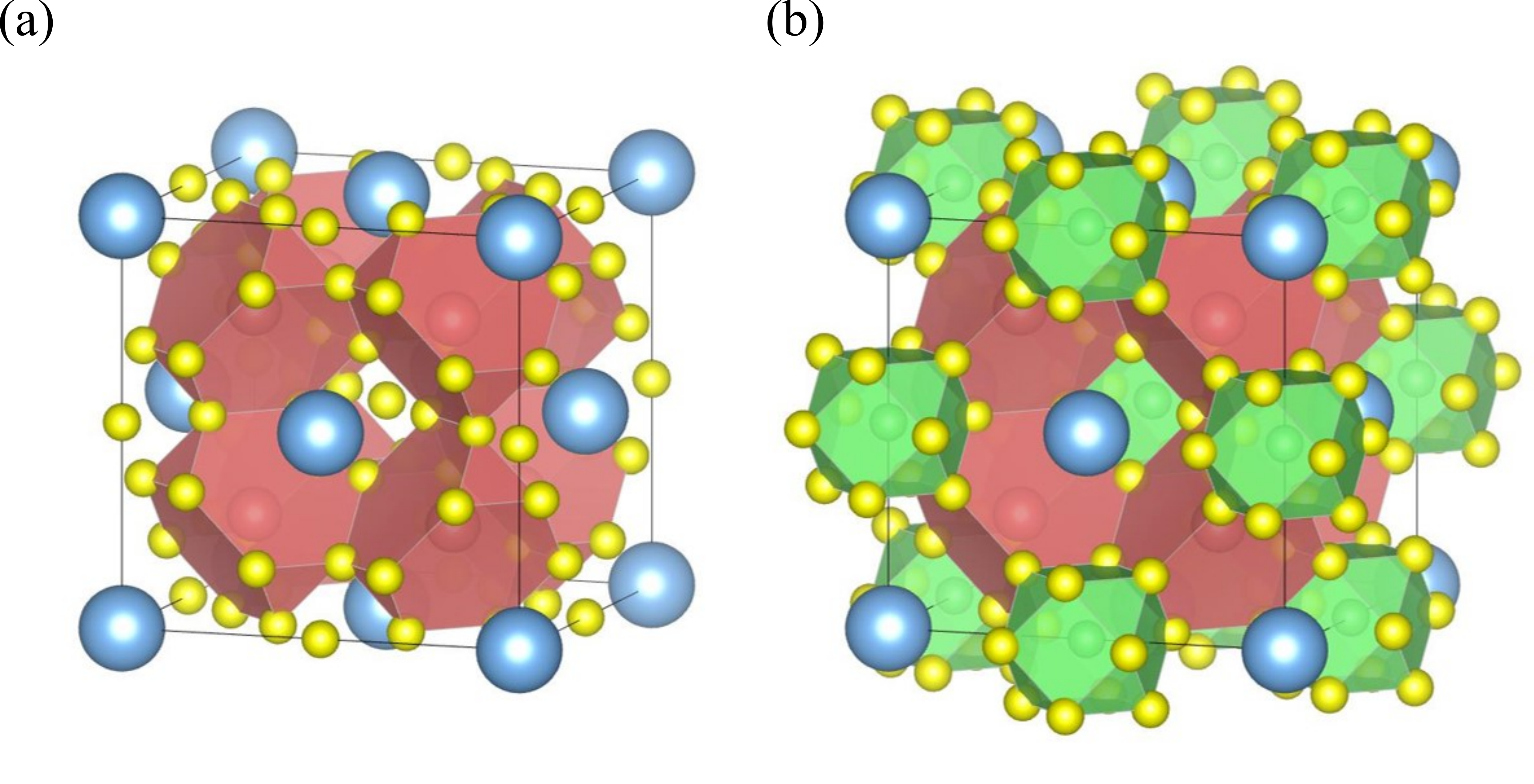}
\caption{In all figures of the paper, the small, medium, and large spheres are represented by yellow, red, and blue balls, respectively, and the green balls correspond to the fourth spheres, named semi-small sphere. 
(a) The (13-2-1) structure~\cite{PhysRevE.104.024101, koshoji2021diverse} with the $Fm \bar{3}m$ symmetry, which is the PDTSP at the several radius ratios including $0.44:0.64:1.00$. (b) The (12-1-2-1) structure with the $Fm \bar{3}m$ symmetry, where one small sphere placed at the center of an octahedral site is substituted by a semi-small sphere. 
Note that the color of polyhedron is the same as that of the center sphere.
All the figures are drawn by VESTA~\cite{Momma:db5098}.
}
\label{fig:13-2-1-prototype}
\end{figure}

\begin{figure}
\centering
\includegraphics[width=1.0\columnwidth]{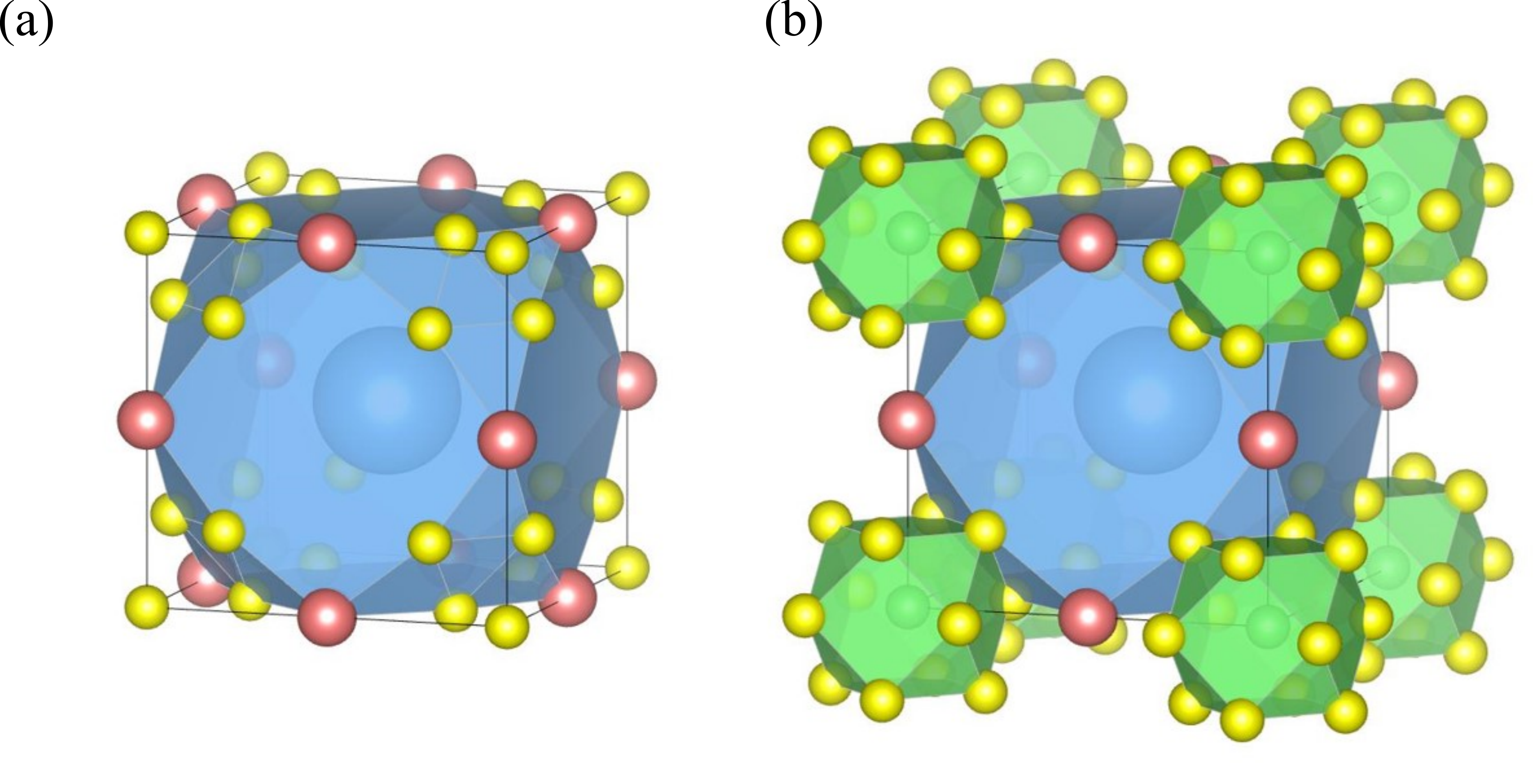}
\caption{(a) The (13-3-1) structure with the $Pm \bar{3}m$ symmetry, which is the PDTSP at the several radius ratios including $0.30:0.40:1.00$ (b) The (12-1-3-1) structure with the $Pm \bar{3}m$ symmetry, where 
one small sphere placed at the vertex of the unit cell is substituted by a semi-small sphere.}
\label{fig:13-3-1-prototype}
\end{figure}

\begin{figure}
\centering
\includegraphics[width=0.9\columnwidth]{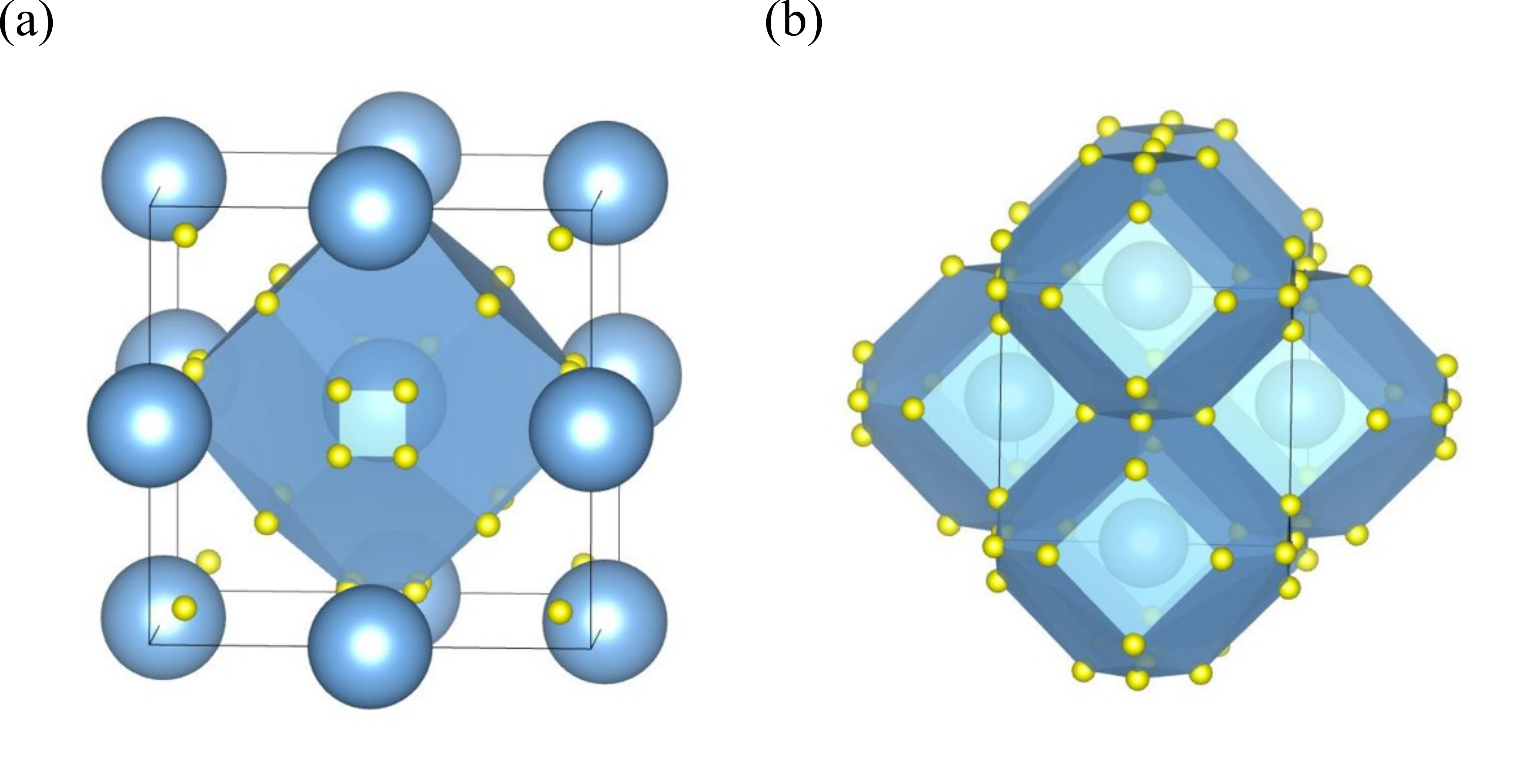}
\caption{Two PDBSPs that correspond to the crystals of clathrate hydrides synthesized under high pressure with high-$T_c$ superconductivity. (a) The $\mathrm{XY}_{10}$ structure~\cite{PhysRevLett.107.125501, PhysRevE.85.021130, PhysRevE.103.023307}, which corresponds to the crystal of $\mathrm{LaH}_{10}$~\cite{doi:10.1002/anie.201709970, Drozdov2019, PhysRevLett.122.027001, doi:10.1063/1.5128736}. (b) The (6-1) structure~\cite{PhysRevLett.107.125501, PhysRevE.85.021130, PhysRevE.103.023307}, which corresponds to the crystal of $\mathrm{YH}_{6}$~\cite{https://doi.org/10.1002/adma.202006832}.}
\label{fig:binary-superhydrides}
\end{figure}

A structural feature similar to the hydrogen sublattice in the clathrate hydrides can be found in some of the putative densest ternary sphere packings (PDTSPs), for example, small spheres in the (13-2-1) and (13-3-1) structures shown in Figs.~\ref{fig:13-2-1-prototype}(a) and \ref{fig:13-3-1-prototype}(a) have cage structures enclosing the medium and large spheres~\cite{PhysRevE.104.024101, koshoji2021diverse}. Note that we name a PDTSP consisting of $l$ small, $m$ medium, and $n$ large spheres per unit cell as ($l$-$m$-$n$) structure. The structural similarity indicates that the two PDTSPs can be regarded as the structural prototypes for clathrate hydrides. In general, the reduction of distances between atoms caused by the high pressure weakens the directional orientation of the bonds due to the strong repulsive force by the Pauli exclusion principle~\cite{doi:10.1021/jacs.9b02634}. So, it is natural to expect that hard-sphere models can help computational discovery of materials under high pressure. In fact, the previous study shows that considerable crystals can be derived from the putative densest binary sphere packings (PDBSPs)~\cite{PhysRevE.103.023307}, for example, the crystal of $\mathrm{LaH}_{10}$ is isotypic to the $\mathrm{XY}_{10}$ structure shown in Fig.~\ref{fig:binary-superhydrides}(a), and similarly the crystal of $\mathrm{YH}_{6}$ is isotypic to the (6-1) structure shown in Fig.~\ref{fig:binary-superhydrides}(b). Furthermore, Zhang \textit{et al.}~\cite{PhysRevLett.128.047001} predicted $\mathrm{LaBeH}_8$, where the tetrahedral site is occupied by a tetrahedron consisting of hydrogen atoms, 
and we see a similar local structure in the (10-4-1) structure that is one of the PDTSPs~\cite{PhysRevE.104.024101}. It is worth pointing out that in general, the PDTSPs tend not to have high symmetries~\cite{PhysRevE.103.023307, PhysRevE.104.024101, koshoji2021diverse}, but the (13-2-1) and (13-3-1) structures have high symmetries of the $Fm \bar{3}m$ and $Pm \bar{3}m$, respectively, if the small structural distortions are corrected.

In this study, we exhaustively search quaternary hydrides based on the (13-2-1) and (13-3-1) PDTSP structures, 
which are used to derive the quaternary structural prototypes by substituting one small sphere in a cluster 
consisting of 13 small spheres with a fourth sphere. For each PDTSP, 36652 kinds of compounds are 
systematically generated by substituting the small spheres with hydrogen atoms, the medium, large, and fourth spheres 
with metallic atoms. Thus, we have 73304 candidate compounds of hydrides in total for the exhaustive search. 
We screen unstable hydrides for the candidates through geometrical optimizations, 
molecular dynamics simulations, and phonon calculations, and identify 23 hydrides
with static and dynamic stability,  
including $\mathrm{H}_{12} \mathrm{Sc} \mathrm{Y}_{2} \mathrm{La}$ 
and $\mathrm{H}_{12} \mathrm{Ti} \mathrm{Ni}_{3} \mathrm{Ba}$.

The paper is organized as follows: Section II describes the conditions for the exhaustive search; 
Sec. III shows the computational results. In Sec. IV, we summarize this study.

\section{Computational details}
\label{sec:method}


\begin{table*}
\caption{The list of substitution atoms for spheres, where the symbol E corresponds to empty.}
\label{table:substitution-chemical-compositions}
\begin{ruledtabular}
\begin{tabular}{cccc}
 & sphere size & elements & \\ \hline
 & small & H & \\
 & semi-small & E, H, Li, Be, Na, Mg, Al, K, Ca, Sc, Ti, V, Cr, Mn, Fe, Co, Ni, Cu, Zn & \\
 & medium & From Li to Hg (except for B to Ne, Si to Ar, Br, Kr, Xe, Ce to Lu, Pt) & \\
 & large & From K to Hg (except for Br, Kr, Xe, Ce to Lu, Pt) & \\
\end{tabular}
\end{ruledtabular}
\end{table*}

\subsection{\textit{Ab initio} simulations}
\label{sec:ab-initio-simulation}

We use the OpenMX code~\cite{PhysRevB.67.155108, PhysRevB.69.195113, PhysRevB.72.045121, openmx} to perform density functional theory (DFT) calculations based on norm-conserving pseudopotentials and optimized pseudoatomic localized basis functions. The exchange-correlation functional was treated within the generalized gradient approximation by Perdew, Burke, and Ernzerhof (GGA-PBE)~\cite{PhysRevLett.77.3865, PhysRev.140.A1133}. The pseudoatomic orbital basis functions are detailed in Supplemental Material \cite{supplemental_material}. 
In geometrical optimizations, lattice vectors and internal coordinates are simultaneously optimized without any constraint by quasi-Newton methods~\cite{doi:10.1021/j100247a015}. We also perform NVT ensemble molecular dynamics simulations (NVTMD) using the Nos\'e-Hoover method~\cite{doi:10.1063/1.447334, doi:10.1080/00268978400101201, PhysRevA.31.1695}, 
and a constant-temperature and constant-pressure molecular dynamics simulations (NPTMD) 
using the velocity scaling method~\cite{WOODCOCK1971257} and the Parrinello-Rahman method~\cite{PhysRevLett.45.1196}.
In the geometrical optimizations, NVTMD, and NPTMD, the $k$-grids of $n_1 \times n_2 \times n_3$
for the Brillouin zone sampling are determined based on the lengths $| \bm{g}_i |$ of reciprocal 
lattice vectors $\bm{g}_i$ [\AA$^{-1}$] as 
\begin{equation}
n_i =\lceil{3 | \bm{g}_i | \rceil}, \label{eq:k-grids}
\end{equation}
and the regular mesh of $230$ Ry in real space was used for the numerical integration and 
for the solution of the Poisson equation \cite{PhysRevB.72.045121}.

To confirm the dynamic stability, we also use the ALAMODE code~\cite{Tadano_2014} to calculate the phonon dispersion. 
The $k$-grids of $n_1 \times n_2 \times n_3$ are set to be twice as large as that calculated as Eq.~(\ref{eq:k-grids}), 
and the regular mesh of $1000$ Ry in real space was used for the numerical integration and for the solution 
of the Poisson equation. In the calculations of force constants, we choose the displacement of $0.1$ {\AA}, 
and calculate the force constants after the symmetrization of structures. 
For the face-centered cubic system, the force constants are calculated 
using the $2 \times 2 \times 2$ conventional cells, while for the primitive cubic system, they
are calculated using the $3 \times 3 \times 3$ primitive cells. 

All the calculations to determine static and dynamic stability are performed 
under hydrostatic pressure of $10$ GPa. 
In the initial screening process, the spin polarization is not taken into account to reduce the computational cost, 
while the spin polarization is considered in the final screening process to determine static and dynamic stability.
The space groups are determined by the code Spglib~\cite{togo2018textttspglib}.

\subsection{Prototype structures and chemical compositions}
\label{sec:structural-prototypes}

One of our prototype structures is the (13-2-1) PDTSP structure~\cite{PhysRevE.104.024101, koshoji2021diverse}, which has the $Fm \bar{3}m$ symmetry as shown in Fig.~\ref{fig:13-2-1-prototype}(a). The large spheres constitute the FCC structure without contact, and a large sphere is surrounded by 24 small spheres constituting truncated octahedron. The tetrahedral site is occupied by one medium sphere which is surrounded by 12 small spheres constituting truncated tetrahedron. The octahedral site is occupied by one small sphere which is surrounded by 12 small spheres constituting cuboctahedron. The structure is the PDTSP at the several radius ratios such as $0.44:0.64:1.00$, at which small and medium spheres 
are too large to be placed in the tetrahedral and octahedral sites if large spheres comprising the FCC structure contact with each other. Thus, the (13-2-1) structure is not a trivial PDTSP. To search the quaternary hydrides, we substitute a small sphere at the center of octahedral site with a fourth sphere named semi-small sphere. We call the quaternary structure shown in Fig.~\ref{fig:13-2-1-prototype}(b) as the (12-1-2-1) structure. This structure also has the $Fm \bar{3}m$ symmetry. The small spheres in the structure enclose the semi-small, medium, and large spheres. It should be noted that the cages of small spheres surrounding large spheres is isotypic to the cages of hydrogen atoms in $\mathrm{LaYH}_{12}$ and $\mathrm{LaY}_3 \mathrm{H}_{24}$~\cite{doi:10.1021/acs.chemmater.1c02371}.

The second prototype structure in the study is the (13-3-1) structure~\cite{PhysRevE.104.024101, koshoji2021diverse}, 
which has the $Pm \bar{3}m$ symmetry as shown in Fig.~\ref{fig:13-3-1-prototype}(a). The structure, which is the PDTSP at the several radius ratios such as $0.30:0.40:1.00$, can be derived by substituting an atom at the vertex of the 
unit cell of the perovskite structure with a cluster of 13 small spheres. 
To search the quaternary hydrides, 
we substitute a small sphere at the center of the cluster of 13 small spheres with a semi-small sphere. 
We call the quaternary structure shown in Fig.~\ref{fig:13-3-1-prototype}(b) as the (12-1-3-1) structure. 
The structure also has the $Pm \bar{3}m$ symmetry. The small spheres in the structure enclose the semi-small, 
medium, and large spheres.

Table \ref{table:substitution-chemical-compositions} shows the list of substitution atoms for spheres in the (12-1-2-1) and (12-1-3-1) structures. Small spheres are substituted with only hydrogen atoms. The number of chemical compositions for each structural prototype is 36652. The initial lattice size of each hydride is determined based on the covalent radii~\cite{B801115J} as follows: we estimate how the lattice must be enlarged for each atom to reduce the overlaps, defined by the covalent radii, to zero with near atoms, and determine the size of the initial lattice by calculating the average of them to expand the lattice moderately.

\subsection{Screening of unstable hydrides}

First, we geometrically optimize the initial structures of the candidates under $10$ GPa. Next, to exclude the unstable hydrides with small computations, we execute the reoptimizations with adding slight structural distortions. Finally, we perform the NVTMD under $10$ GPa with time step of $0.5$ femtosecond (fs) for $1.0$ picosecond at $200$ K in the $1 \times 1 \times 1$ unit cell. If a hydride transforms a disordered structure, we regard that the hydride is dynamically unstable. In this screening process, we do not take account of the spin polarization to reduce the computational cost.

\subsection{Standard enthalpies of formations}

The standard enthalpy of formation (SEF) is defined by the enthalpy of a material minus the sum of those of elemental materials. If the SEF is positive, the material is less probable to be synthesized. When we calculate the enthalpy of an elemental 
material under 10 GPa, the crystal structure is set to be isotypic to that under the 0 GPa, i.e., only the lattice size
is optimized under a given pressure. We note that the crystal of lanthanum metal is set to be the FCC structure and 
the spin polarization is taken into account in all the calculations.

\subsection{Confirmation of structural stabilities}

The structural stability is confirmed by phonon dispersion and NPTMD. 
If a material has no imaginary frequency mode of phonon, we regard that the structure is dynamically stable. 
The NPTMD is performed under $10$ GPa with time step of $0.5$ fs for $0.5$ picosecond at $200$ K 
in the $2 \times 2 \times 2$ unit cell. Since these calculations need large computational costs, 
we perform them for some of hydrides which have lower SEFs and have no spin polarization.

\section{Results}
\label{sec:results}

\begin{figure}
\centering
\includegraphics[width=1.0\columnwidth]{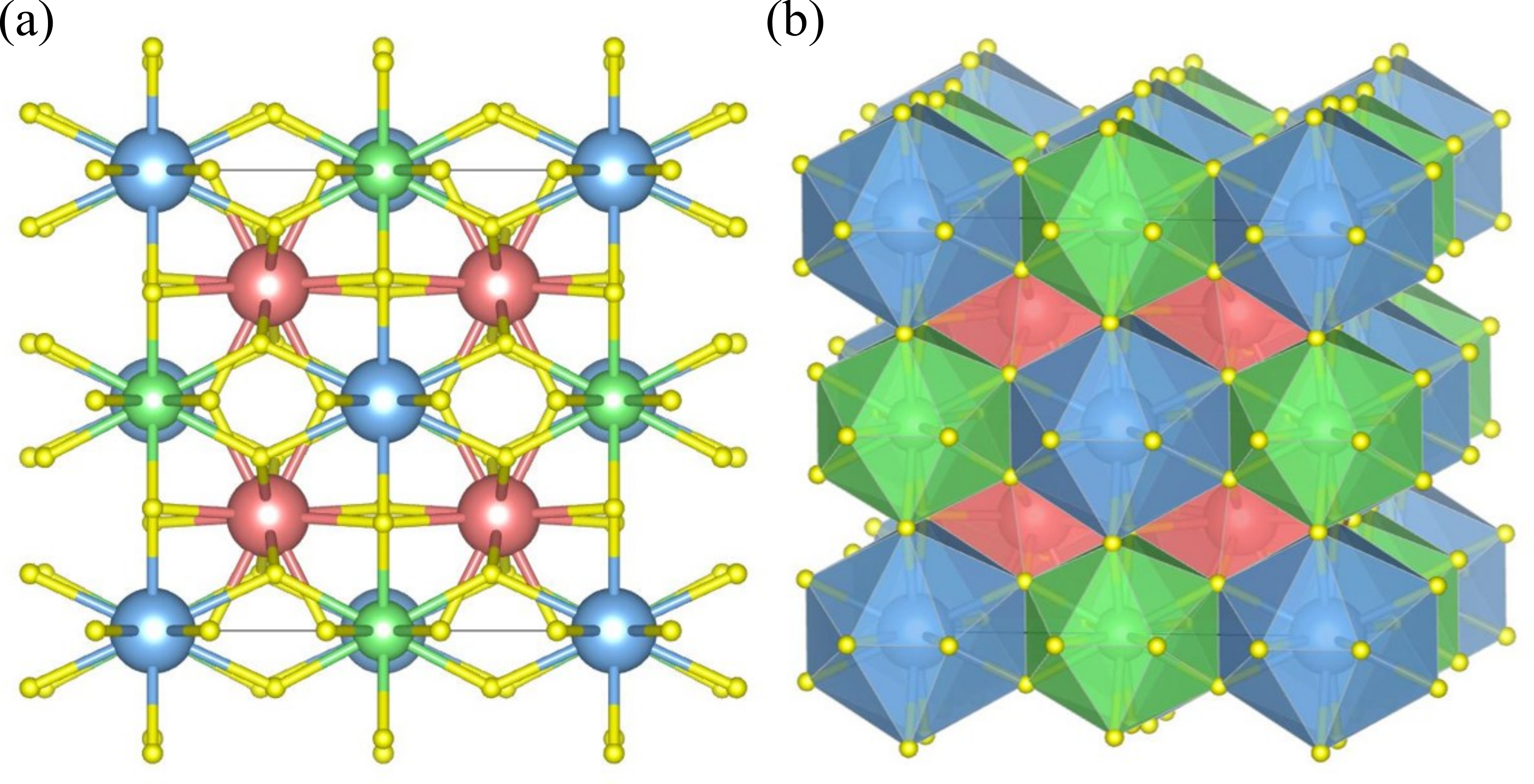}
\caption{The conventional cell of the $\mathrm{H}_{12} \mathrm{ScY}_2 \mathrm{La}$. This structure is the variant structure of the (12-1-2-1) structure named (12-1-2-1)$_V$ structure, and it has the $Fm\bar{3}$ symmetry. The yellow, green, red, and blue balls correspond to the hydrogen, scandium, yttrium, and lanthanum atoms, respectively. (a) The crystal is shown by balls and sticks. (b) The crystal is shown by polyhedrons consisting of hydrogen atoms.}
\label{fig:H_12_Sc_1_Y_2_La_1}
\end{figure}
\begin{figure}
\centering
\includegraphics[width=1.0\columnwidth]{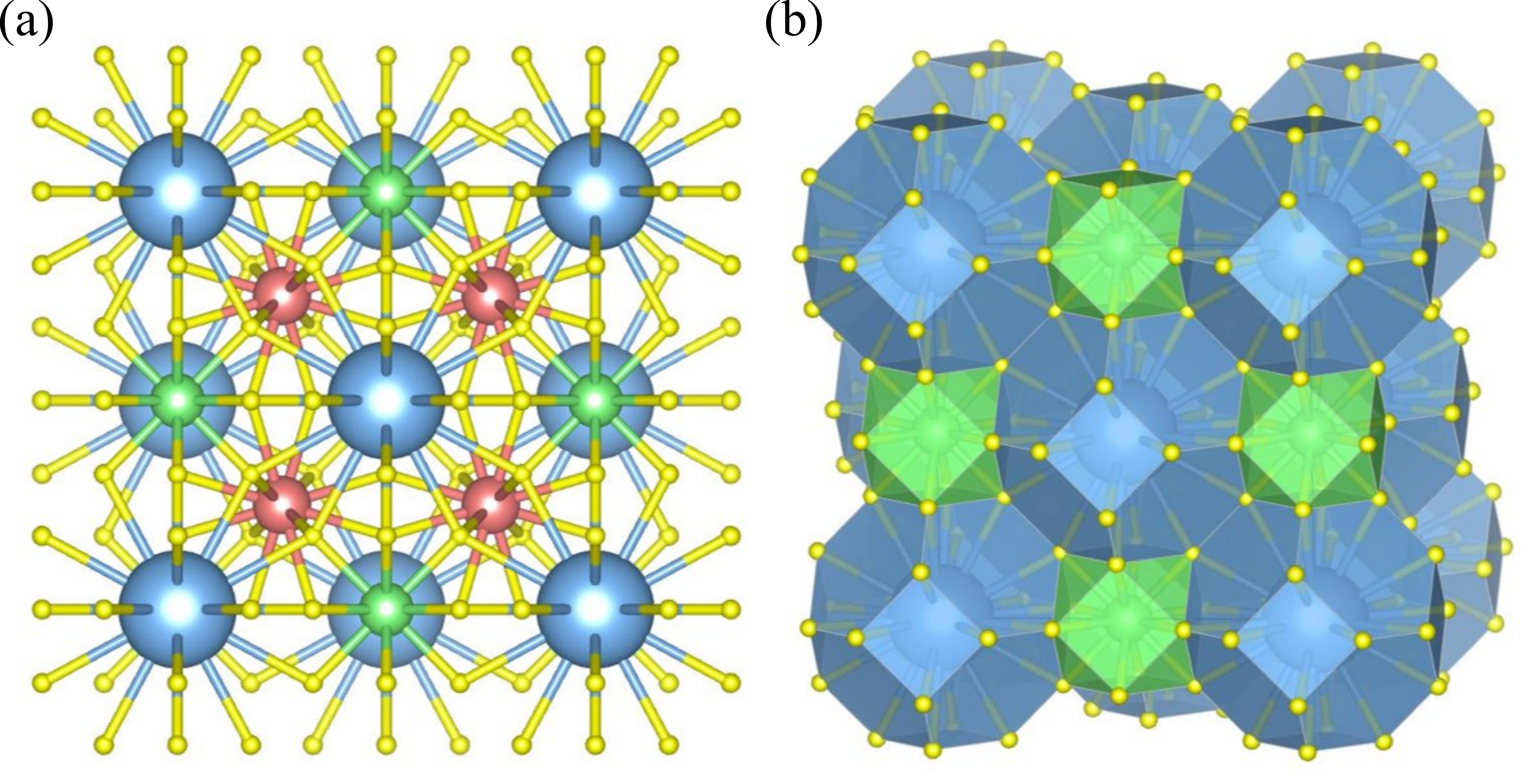}
\caption{The conventional cell of the $\mathrm{H}_{12} \mathrm{TiSc}_2 \mathrm{Cs}$. The yellow, green, red, and blue balls correspond to the hydrogen, titanium, scandium, and cesium atoms, respectively.}
\label{fig:H_12_Ti_1_Sc_2_Cs_1}
\end{figure}
\begin{figure}
\centering
\includegraphics[width=1.0\columnwidth]{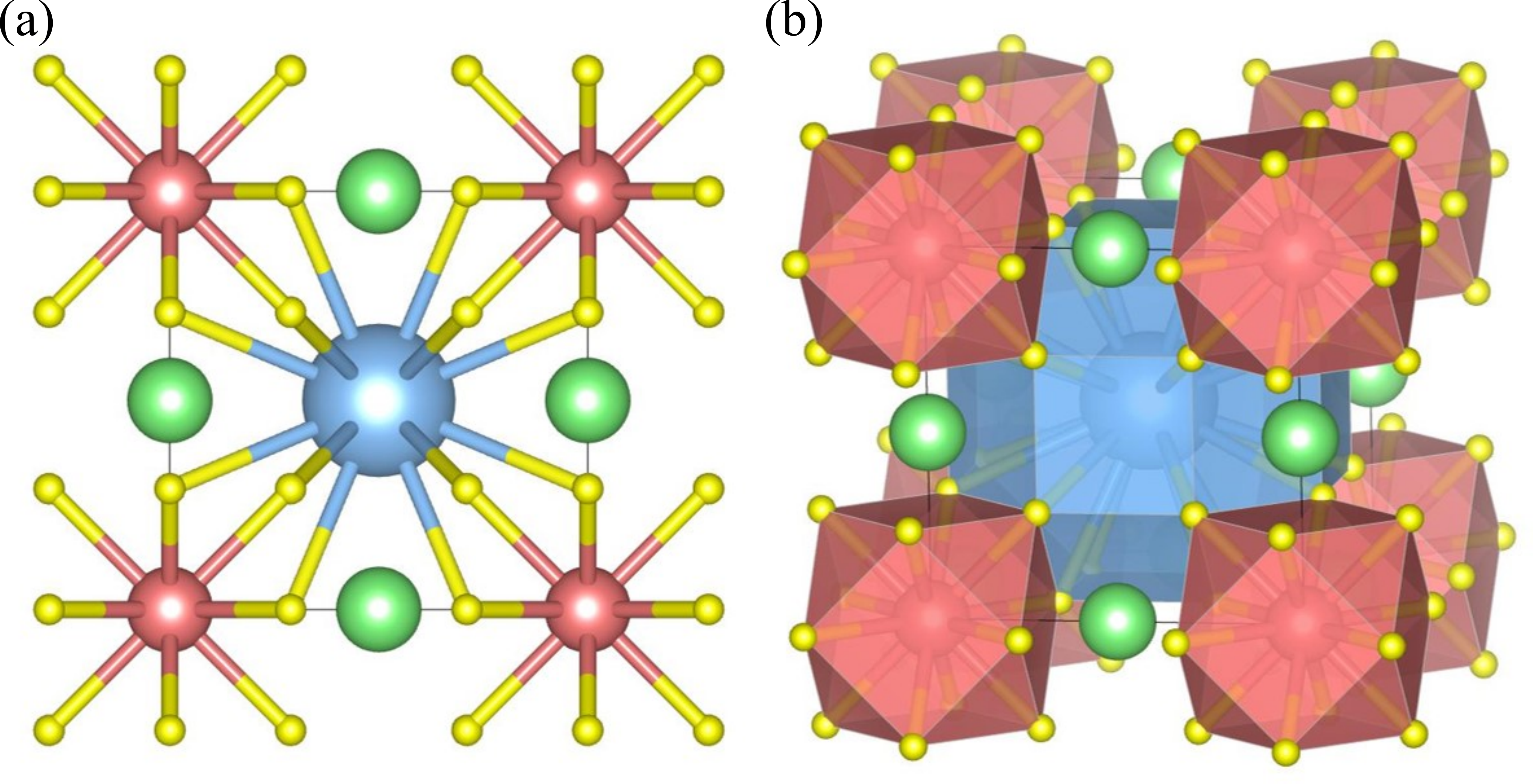}
\caption{The primitive cell of the $\mathrm{H}_{12} \mathrm{TiNi}_3 \mathrm{Ba}$. The yellow, green, red, and blue balls correspond to the hydrogen, titanium, nickel, and barium atoms, respectively.}
\label{fig:H_12_Ti_1_Ni_3_Ba_1}
\end{figure}

\begin{table}
\caption{The 16 kinds of (12-1-2-1)$_V$-type hydrides which have the SEFs less than $-10.0$ eV/f.u. 
         and no spin polarization, and show the dynamic stability confirmed by both the phonon 
         calculation and the NPTMD, except for $\mathrm{H}_{12} \mathrm{Sc} \mathrm{La}_{2} \mathrm{Ba}$
         marked by $*$ for which only the phonon calculation supports the dynamic stability.}
\label{table:202_12-1-2-1_materials_nospin}
\begin{ruledtabular}
\begin{tabular}{cccc}
 & SEFs (eV/f.u.) & Material names & \\ \hline
 & $-12.20$ & $\mathrm{H}_{12} \mathrm{Sc} \mathrm{Y}_{2} \mathrm{La}$ & \\
 & $-11.87$ & $\mathrm{H}_{12} \mathrm{Sc} \mathrm{Sc}_{2} \mathrm{Y}$ & \\
 & $-11.60$ & $\mathrm{H}_{12} \mathrm{Sc} \mathrm{Y}_{2} \mathrm{Zr}$ & \\
 & $-11.58$ & $\mathrm{H}_{12} \mathrm{Sc} \mathrm{Y}_{2} \mathrm{Hf}$ & \\
 & $-11.45$ & $\mathrm{H}_{12} \mathrm{Sc} \mathrm{Sc}_{2} \mathrm{Hf}$ & \\
 & $-11.12$ & $\mathrm{H}_{12} \mathrm{Sc} \mathrm{Y}_{2} \mathrm{Ca}$ & \\
 & $-11.05$ & $\mathrm{H}_{12} \mathrm{Sc} \mathrm{Y}_{2} \mathrm{Sr}$ & \\
 & $-10.82$ & $\mathrm{H}_{12} \mathrm{Ti} \mathrm{Y}_{2} \mathrm{Ca}$ & \\
 & $-10.78$ & $\mathrm{H}_{12} \mathrm{Ti} \mathrm{Y}_{2} \mathrm{Sr}$ & \\
 & $-10.49$ & $\mathrm{H}_{12} \mathrm{Sc} \mathrm{La}_{2} \mathrm{Sr}$ & \\
 & $-10.26$ & $\mathrm{H}_{12} \mathrm{Sc} \mathrm{La}_{2} \mathrm{Ba}^*$ & \\
 & $-10.15$ & $\mathrm{H}_{12} \mathrm{Ca} \mathrm{Zr}_{2} \mathrm{Zr}$ & \\
 & $-10.15$ & $\mathrm{H}_{12} \mathrm{Ti} \mathrm{La}_{2} \mathrm{Sr}$ & \\
 & $-10.13$ & $\mathrm{H}_{12} \mathrm{Ti} \mathrm{Sc}_{2} \mathrm{Zr}$ & \\
 & $-10.12$ & $\mathrm{H}_{12} \mathrm{Ti} \mathrm{Y}_{2} \mathrm{Ba}$ & \\
 & $-10.00$ & $\mathrm{H}_{12} \mathrm{Ti} \mathrm{La}_{2} \mathrm{Ba}$ &
\end{tabular}
\end{ruledtabular}
\end{table}
\begin{table}
\caption{The six kinds of (12-1-3-1)-type hydrides which have the SEFs less than $-5.0$ eV/f.u. and no spin polarization, and show the dynamic stability confirmed by both the phonon calculation and the NPTMD, except for $\mathrm{H}_{12} \mathrm{Sc} \mathrm{Ni}_{3} \mathrm{La}$ marked by $*$ for which only the phonon calculation supports the dynamic stability.}
\label{table:12-1-3-1_materials}
\begin{ruledtabular}
\begin{tabular}{cccc}
 & SEFs (eV/f.u.) & Material names & \\ \hline
 & $-6.47$ & $\mathrm{H}_{12} \mathrm{Ti} \mathrm{Ni}_{3} \mathrm{Ba}$ & \\
 & $-5.82$ & $\mathrm{H}_{12} \mathrm{Ti} \mathrm{Ni}_{3} \mathrm{Sr}$ & \\
 & $-5.72$ & $\mathrm{H}_{12} \mathrm{Sc} \mathrm{Ni}_{3} \mathrm{La}^*$ & \\
 & $-5.69$ & $\mathrm{H}_{12} \mathrm{Sc} \mathrm{Pd}_{3} \mathrm{Ba}$ & \\
 & $-5.50$ & $\mathrm{H}_{12} \mathrm{Li} \mathrm{Ru}_{3} \mathrm{Ba}$ & \\
 & $-5.07$ & $\mathrm{H}_{12} \mathrm{Ti} \mathrm{Pd}_{3} \mathrm{Ba}$ &
\end{tabular}
\end{ruledtabular}
\end{table}

\begin{figure}
\centering
\includegraphics[width=0.72 \columnwidth]{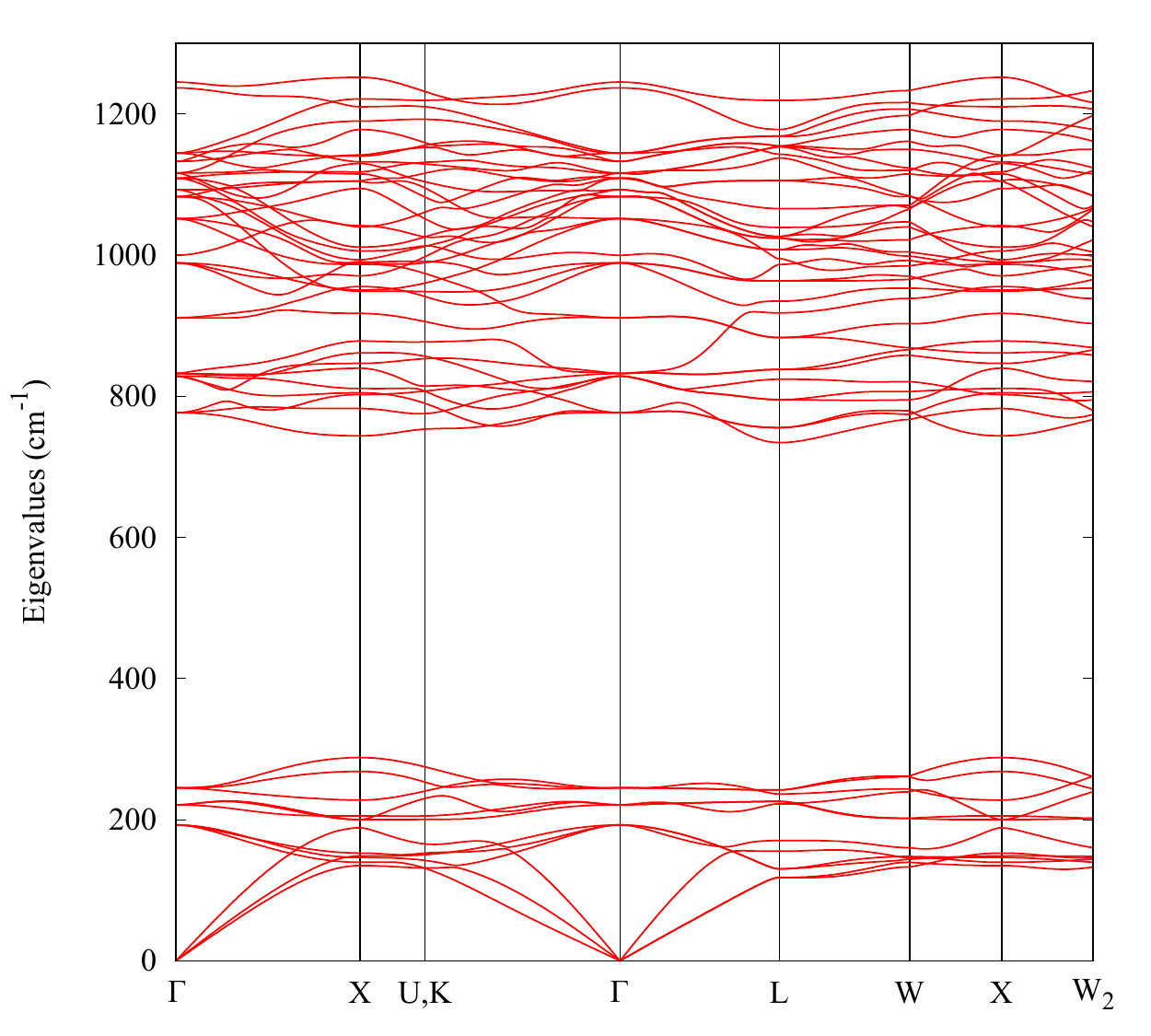}
\caption{The phonon bands of $\mathrm{H}_{12} \mathrm{ScY}_2 \mathrm{Ca}$ at 10 GPa. As represented in $\mathrm{H}_{12} \mathrm{ScY}_2 \mathrm{Ca}$, some of the (12-1-2-1)$_V$-type hydrides have large gaps in phonon bands and high phonon frequency modes.}
\label{fig:H_12_Sc_1_Y_2_Ca_1_pband}
\end{figure}
\begin{figure}
\centering
\includegraphics[width=0.95\columnwidth]{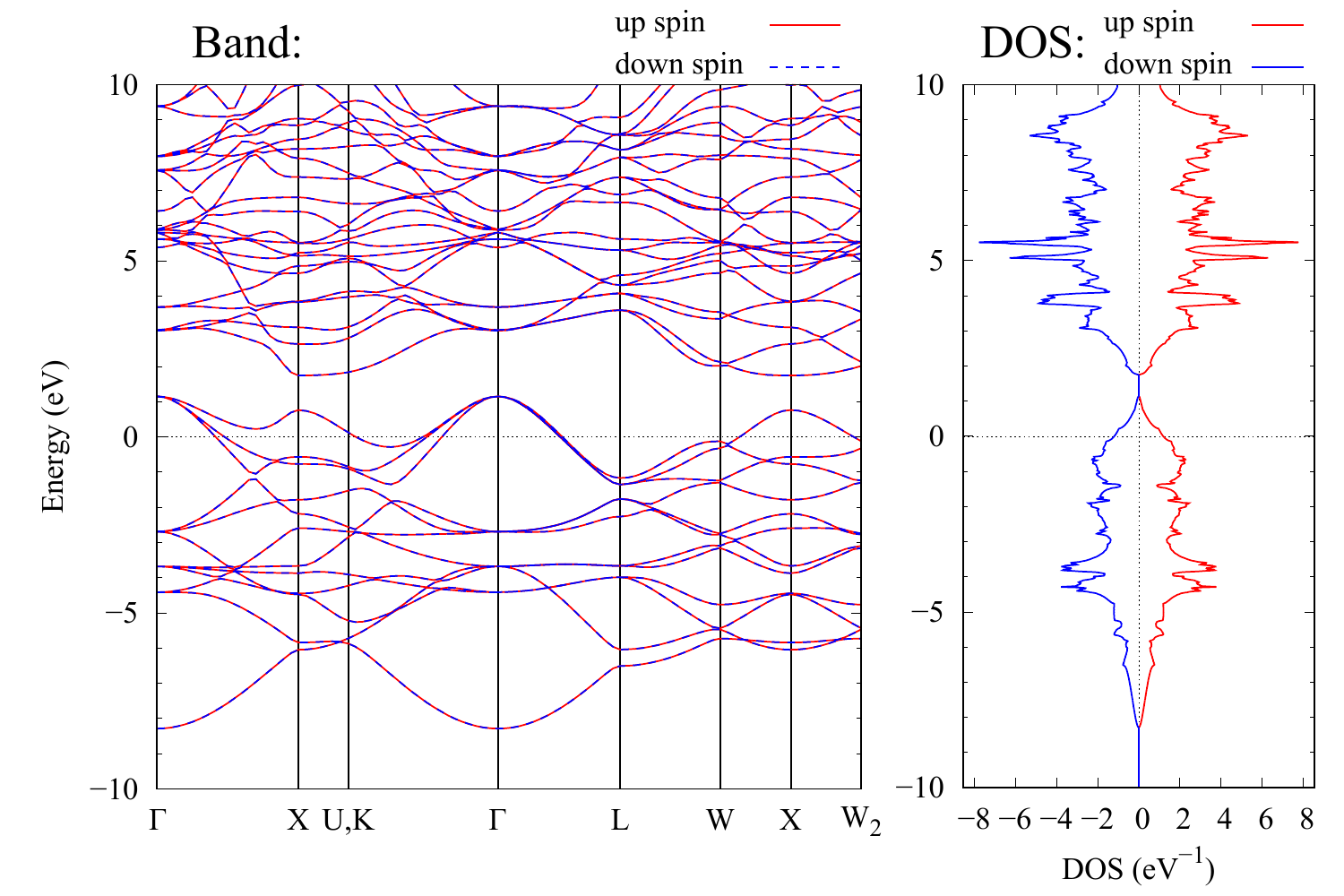}
\caption{The bands and density of state (DOS) of $\mathrm{H}_{12} \mathrm{ScY}_2 \mathrm{Ca}$ at 10 GPa. As represented in $\mathrm{H}_{12} \mathrm{ScY}_2 \mathrm{Ca}$, some of the (12-1-2-1)$_V$-type hydrides have large DOS near the Fermi level.}
\label{fig:H_12_Sc_1_Y_2_Ca_1_band_dos}
\end{figure}

After generating 73304 candidate compounds derived from the (12-1-2-1) and (12-1-3-1) structures, we exclude unstable hydrides by geometrical optimizations and NVTMDs. Among the hydrides that keep symmetric structures in the process, we calculate the phonon dispersions and perform the NPTMDs for 28 hydrides which have the lowest SFEs and exhibit no spin polarization in order to confirm the dynamic stability.

In the NVTMDs, we find that almost all the candidates generated from the (12-1-2-1) structure do not prefer their original structure but another one with the $Fm\bar{3}$ symmetry named (12-1-2-1)$_V$ structure shown in Fig.~\ref{fig:H_12_Sc_1_Y_2_La_1} \cite{MgGeH6}. 
Among compounds with the (12-1-2-1)$_V$ structure, we calculate the phonon dispersions for 18 hydrides which have the SEFs 
less than $-10.0$ eV/f.u. and exhibits no spin polarization even in the spin polarized calculations. 
As a result, we confirm that 16 hydrides listed in Table \ref{table:202_12-1-2-1_materials_nospin} are dynamically stable 
as shown in Figs. 1 and 2 of Supplemental Material \cite{supplemental_material}. 
Tables I and II of Supplemental Material \cite{supplemental_material} list the (12-1-2-1)$_V$-type hydrides 
having the SEFs in between $-10.0$ and $-5.0$, and $-5.0$ and $-1.3$ eV/f.u., respectively, which show the dynamic 
stability in the NVTMDs and have no spin polarization. 
In addition, Table III of Supplemental Material \cite{supplemental_material} lists the (12-1-2-1)$_V$-type hydrides 
which also show the dynamic stability in the NVTMDs with {\it spin} polarization. 
The NPTMDs under 10 GPa also support the results of phonon calculations except for
$\mathrm{H}_{12} \mathrm{Sc} \mathrm{La}_{2} \mathrm{Ba}$, marked by $*$ 
in Table \ref{table:202_12-1-2-1_materials_nospin}, exhibiting the dynamic instability in the NPTMD.
As an example of the (12-1-2-1)$_V$ structure, we show the phonon dispersion and electronic band structure 
of $\mathrm{H}_{12} \mathrm{ScY}_2 \mathrm{Ca}$ in Figs.~\ref{fig:H_12_Sc_1_Y_2_Ca_1_pband} and 
\ref{fig:H_12_Sc_1_Y_2_Ca_1_band_dos}, respectively. 
We see that $\mathrm{H}_{12} \mathrm{ScY}_2 \mathrm{Ca}$ has high phonon frequency modes and 
a large density of states (DOS) near the Fermi level. 
The phonon dispersions and band structures of the other 15 hydrides with the (12-1-2-1)$_V$ structure 
are shown in Figs.~1-4 of Supplemental Material \cite{supplemental_material}. 

In the NVTMDs we see that the original (12-1-2-1) structure is preferred by a few candidate compounds. 
To further check the dynamic stability of the compounds with the (12-1-2-1) structure, 
we calculate the phonon dispersions of the four hydrides that have the lowest SFEs in them 
except for those having spin polarizations and relatively large structural distortions. 
The result indicates that only $\mathrm{H}_{12} \mathrm{TiSc}_2 \mathrm{Cs}$ shown in Fig.~\ref{fig:H_12_Ti_1_Sc_2_Cs_1} is dynamically stable. The phonon dispersion and electronic band structure are shown in Figs.~5 and 6 
of Supplemental Material \cite{supplemental_material}, respectively. 
Although the phonon calculations show the dynamic stability of only $\mathrm{H}_{12} \mathrm{TiSc}_2 \mathrm{Cs}$ 
in the compounds with the original (12-1-2-1) structure, we note that the NPTMDs suggest the dynamic stability 
of not only $\mathrm{H}_{12} \mathrm{TiSc}_2 \mathrm{Cs}$, 
but also $\mathrm{H}_{12} \mathrm{VSc}_2 \mathrm{Cs}$, which have the SEFs of $-7.70$ eV/f.u. and $-6.31$ eV/f.u., 
respectively.

On the other hand, the (12-1-3-1) structure is preferred in NVTMDs by a considerable number of 
hydrides including $\mathrm{H}_{12} \mathrm{Ti} \mathrm{Ni}_{3} \mathrm{Ba}$ as shown 
in Fig.~\ref{fig:H_12_Ti_1_Ni_3_Ba_1}. 
Among hydrides with the (12-1-3-1) structure, we calculate the phonon dispersions of six hydrides 
which have the SEFs less than $-5.0$ eV/f.u. and exhibit no spin polarization even in the spin polarized calculations.
As a result, all of them, which are listed in Table \ref{table:12-1-3-1_materials}, show the dynamic stability 
as shown in Fig.~7 of Supplemental Material \cite{supplemental_material}. 
The NPTMDs under 10 GPa also support the results of phonon calculations except for
$\mathrm{H}_{12} \mathrm{Sc} \mathrm{Ni}_{3} \mathrm{La}$, marked by $*$ 
in Table \ref{table:12-1-3-1_materials}, exhibiting the dynamic instability in the NPTMD.
The electronic band structures of the (12-1-3-1)-type hydrides listed in Table \ref{table:12-1-3-1_materials}
are shown in Fig.~8 of Supplemental Material \cite{supplemental_material}, which show that all the hydrides are metal.
In addition to Table III, Table IV of Supplemental Material \cite{supplemental_material} list 
the (12-1-3-1)-type hydrides which show the dynamic stability in the NVTMDs, and have the SEFs more than $-5.0$ eV/f.u. 
and no spin polarization. Also, Table V of Supplemental Material \cite{supplemental_material} lists 
the (12-1-3-1)-type hydrides which also show the dynamic stability in NVTMDs but exhibit {\it spin} polarizations. 

As listed in Table \ref{table:substitution-chemical-compositions}, we also substitute semi-small spheres 
with hydrogen atoms or empty, however, our calculations suggest that such ternary hydrides are 
less probable to be synthesized at least under $10$ GPa.

\section{conclusions}
\label{sec:conclusion}

Some of the 60 kinds of PDTSPs have high symmetries and unique structural properties~\cite{PhysRevE.104.024101, koshoji2021diverse}. In this study, we focus on the (13-2-1) and (13-3-1) PDTSP structures, 
which can be utilized to derive the quaternary structural prototypes named (12-1-2-1) and (12-1-3-1) structures 
by substituting one small sphere in a cluster consisting of 13 small spheres with a fourth sphere. 
Substituting the small spheres in the two prototypes with hydrogen atoms gives unconventional hydrogen sublattices 
in cubic systems. Based on the two prototypes, we exhaustively search the quaternary metal hydrides 
from 73304 kinds of the candidates. First, we exclude unstable hydrides by geometrical optimizations and NVTMDs. 
In this process, almost all the candidates generated from the (12-1-2-1) structure do not prefer their original structure 
but another one with the $Fm\bar{3}$ symmetry named (12-1-2-1)$_V$ structure. 
Second, we calculate the SFEs including spin polarization to estimate the static stability. 
We find that the (12-1-2-1)$_V$-type hydrides tend to have lower SFEs than the other two type hydrides. 
Third, to confirm the dynamic stability of hydrides that have lower SEFs and no spin polarization, 
we calculate the phonon dispersions of 4 kinds of (12-1-2-1)-type hydrides, 18 kinds of (12-1-2-1)$_V$-type hydrides, 
and 6 kinds of (12-1-3-1)-type hydrides. 
As a result, one $\mathrm{H}_{12} \mathrm{TiSc}_2 \mathrm{Cs}$ among the (12-1-2-1)-type hydrides, 
16 hydrides among the (12-1-2-1)$_V$-type hydrides, and all of the (12-1-3-1)-type hydrides show the dynamic stability. 
Some of them have large gaps in phonon bands, high phonon frequency modes, and large DOS near the Fermi level. 
From the similarities with several cubic hydrides such as $\mathrm{YH}_6$ and $\mathrm{LaH}_{10}$, 
which are high-temperature SCs under high pressures~\cite{PhysRevLett.119.107001, Liu6990}, 
one can expect that their hydrogen sublattices is predominantly responsible for superconductivity with 
a high critical temperature. Finally, the NPTMDs for the 28 hydrides support the dynamic stability 
in most cases as well as the results of the phonon calculations. 

Not only the (13-2-1) and (13-3-1) structures but also some of the PDTSPs have high symmetries and unique structures 
that are difficult to design based on the perspectives of chemical bonds and/or local polyhedrons found in transition 
elements. We expect that such structural prototypes can help discover novel crystal structures, 
especially under high pressure. In addition, some of the local structures in the PDTSPs might be useful 
as the building blocks in crystals to design unknown structural prototypes~\cite{koshoji2021diverse, PhysRevLett.128.047001}.

We do not prove that the predicted hydrides can be synthesized in experience, since we do not show 
whether or not they are more stable than any other phases such as decomposition into unary, binary, 
and ternary phases. However, the 23 candidate compounds screened by the exhaustive search 
for 73304 hydrides provide a guideline to narrow down the search space for trials in the experimental synthesis of quaternary metal hydrides.

\begin{acknowledgments}

R. K. is financially supported by the Quantum Science and Technology Fellowship Program (Q-STEP) that is the University Fellowship program for Science and Technology Innovations, and the Grant-in-Aid for JSPS Research Fellow.
The computations in this study have been done using the facilities (supercomputer Ohtaka) of
the Supercomputer Center, the Institute for Solid State Physics, the University of Tokyo
and also using the Fujitsu PRIMERGY CX400M1/CX2550M5 (Oakbridge-CX) at the Information Technology Center, The University of Tokyo.
The authors would like to thank Masanobu Miyata and Mikio Koyano for kindly advising how to use the ALAMODE code to calculate phonon properties.

\end{acknowledgments}

\bibliographystyle{apsrev4-2}
\bibliography{submit}

\begin{thebibliography}{73}%
\makeatletter
\providecommand \@ifxundefined [1]{%
 \@ifx{#1\undefined}
}%
\providecommand \@ifnum [1]{%
 \ifnum #1\expandafter \@firstoftwo
 \else \expandafter \@secondoftwo
 \fi
}%
\providecommand \@ifx [1]{%
 \ifx #1\expandafter \@firstoftwo
 \else \expandafter \@secondoftwo
 \fi
}%
\providecommand \natexlab [1]{#1}%
\providecommand \enquote  [1]{``#1''}%
\providecommand \bibnamefont  [1]{#1}%
\providecommand \bibfnamefont [1]{#1}%
\providecommand \citenamefont [1]{#1}%
\providecommand \href@noop [0]{\@secondoftwo}%
\providecommand \href [0]{\begingroup \@sanitize@url \@href}%
\providecommand \@href[1]{\@@startlink{#1}\@@href}%
\providecommand \@@href[1]{\endgroup#1\@@endlink}%
\providecommand \@sanitize@url [0]{\catcode `\\12\catcode `\$12\catcode
  `\&12\catcode `\#12\catcode `\^12\catcode `\_12\catcode `\%12\relax}%
\providecommand \@@startlink[1]{}%
\providecommand \@@endlink[0]{}%
\providecommand \url  [0]{\begingroup\@sanitize@url \@url }%
\providecommand \@url [1]{\endgroup\@href {#1}{\urlprefix }}%
\providecommand \urlprefix  [0]{URL }%
\providecommand \Eprint [0]{\href }%
\providecommand \doibase [0]{https://doi.org/}%
\providecommand \selectlanguage [0]{\@gobble}%
\providecommand \bibinfo  [0]{\@secondoftwo}%
\providecommand \bibfield  [0]{\@secondoftwo}%
\providecommand \translation [1]{[#1]}%
\providecommand \BibitemOpen [0]{}%
\providecommand \bibitemStop [0]{}%
\providecommand \bibitemNoStop [0]{.\EOS\space}%
\providecommand \EOS [0]{\spacefactor3000\relax}%
\providecommand \BibitemShut  [1]{\csname bibitem#1\endcsname}%
\let\auto@bib@innerbib\@empty
\bibitem [{\citenamefont {Oganov}\ and\ \citenamefont
  {Glass}(2006)}]{doi:10.1063/1.2210932}%
  \BibitemOpen
  \bibfield  {author} {\bibinfo {author} {\bibfnamefont {A.~R.}\ \bibnamefont
  {Oganov}}\ and\ \bibinfo {author} {\bibfnamefont {C.~W.}\ \bibnamefont
  {Glass}},\ }\href {https://doi.org/10.1063/1.2210932} {\bibfield  {journal}
  {\bibinfo  {journal} {The Journal of Chemical Physics}\ }\textbf {\bibinfo
  {volume} {124}},\ \bibinfo {pages} {244704} (\bibinfo {year}
  {2006})}\BibitemShut {NoStop}%
\bibitem [{\citenamefont {Oganov}\ \emph {et~al.}(2011)\citenamefont {Oganov},
  \citenamefont {Lyakhov},\ and\ \citenamefont
  {Valle}}]{doi:10.1021/ar1001318}%
  \BibitemOpen
  \bibfield  {author} {\bibinfo {author} {\bibfnamefont {A.~R.}\ \bibnamefont
  {Oganov}}, \bibinfo {author} {\bibfnamefont {A.~O.}\ \bibnamefont
  {Lyakhov}},\ and\ \bibinfo {author} {\bibfnamefont {M.}~\bibnamefont
  {Valle}},\ }\href {https://doi.org/10.1021/ar1001318} {\bibfield  {journal}
  {\bibinfo  {journal} {Accounts of Chemical Research}\ }\textbf {\bibinfo
  {volume} {44}},\ \bibinfo {pages} {227} (\bibinfo {year} {2011})}\BibitemShut
  {NoStop}%
\bibitem [{\citenamefont {Lyakhov}\ \emph {et~al.}(2013)\citenamefont
  {Lyakhov}, \citenamefont {Oganov}, \citenamefont {Stokes},\ and\
  \citenamefont {Zhu}}]{LYAKHOV20131172}%
  \BibitemOpen
  \bibfield  {author} {\bibinfo {author} {\bibfnamefont {A.~O.}\ \bibnamefont
  {Lyakhov}}, \bibinfo {author} {\bibfnamefont {A.~R.}\ \bibnamefont {Oganov}},
  \bibinfo {author} {\bibfnamefont {H.~T.}\ \bibnamefont {Stokes}},\ and\
  \bibinfo {author} {\bibfnamefont {Q.}~\bibnamefont {Zhu}},\ }\href
  {https://doi.org/https://doi.org/10.1016/j.cpc.2012.12.009} {\bibfield
  {journal} {\bibinfo  {journal} {Computer Physics Communications}\ }\textbf
  {\bibinfo {volume} {184}},\ \bibinfo {pages} {1172 } (\bibinfo {year}
  {2013})}\BibitemShut {NoStop}%
\bibitem [{\citenamefont {Oganov}\ \emph {et~al.}(2010)\citenamefont {Oganov},
  \citenamefont {Ma}, \citenamefont {Lyakhov}, \citenamefont {Valle},\ and\
  \citenamefont {Gatti}}]{10.2138/rmg.2010.71.13}%
  \BibitemOpen
  \bibfield  {author} {\bibinfo {author} {\bibfnamefont {A.~R.}\ \bibnamefont
  {Oganov}}, \bibinfo {author} {\bibfnamefont {Y.}~\bibnamefont {Ma}}, \bibinfo
  {author} {\bibfnamefont {A.~O.}\ \bibnamefont {Lyakhov}}, \bibinfo {author}
  {\bibfnamefont {M.}~\bibnamefont {Valle}},\ and\ \bibinfo {author}
  {\bibfnamefont {C.}~\bibnamefont {Gatti}},\ }\href
  {https://doi.org/10.2138/rmg.2010.71.13} {\bibfield  {journal} {\bibinfo
  {journal} {Reviews in Mineralogy and Geochemistry}\ }\textbf {\bibinfo
  {volume} {71}},\ \bibinfo {pages} {271} (\bibinfo {year} {2010})}\BibitemShut
  {NoStop}%
\bibitem [{\citenamefont {Oganov}\ \emph {et~al.}(2019)\citenamefont {Oganov},
  \citenamefont {Saleh},\ and\ \citenamefont
  {Kvashnin}}]{10.1039/9781788010122}%
  \BibitemOpen
  \bibinfo {editor} {\bibfnamefont {A.~R.}\ \bibnamefont {Oganov}}, \bibinfo
  {editor} {\bibfnamefont {G.}~\bibnamefont {Saleh}},\ and\ \bibinfo {editor}
  {\bibfnamefont {A.~G.}\ \bibnamefont {Kvashnin}},\ eds.,\ \href
  {https://doi.org/10.1039/9781788010122} {\emph {\bibinfo {title}
  {Computational Materials Discovery}}}\ (\bibinfo  {publisher} {The Royal
  Society of Chemistry},\ \bibinfo {year} {2019})\BibitemShut {NoStop}%
\bibitem [{\citenamefont {Wang}\ \emph {et~al.}(2010)\citenamefont {Wang},
  \citenamefont {Lv}, \citenamefont {Zhu},\ and\ \citenamefont
  {Ma}}]{PhysRevB.82.094116}%
  \BibitemOpen
  \bibfield  {author} {\bibinfo {author} {\bibfnamefont {Y.}~\bibnamefont
  {Wang}}, \bibinfo {author} {\bibfnamefont {J.}~\bibnamefont {Lv}}, \bibinfo
  {author} {\bibfnamefont {L.}~\bibnamefont {Zhu}},\ and\ \bibinfo {author}
  {\bibfnamefont {Y.}~\bibnamefont {Ma}},\ }\href
  {https://doi.org/10.1103/PhysRevB.82.094116} {\bibfield  {journal} {\bibinfo
  {journal} {Phys. Rev. B}\ }\textbf {\bibinfo {volume} {82}},\ \bibinfo
  {pages} {094116} (\bibinfo {year} {2010})}\BibitemShut {NoStop}%
\bibitem [{\citenamefont {Wang}\ \emph {et~al.}(2012)\citenamefont {Wang},
  \citenamefont {Lv}, \citenamefont {Zhu},\ and\ \citenamefont
  {Ma}}]{WANG20122063}%
  \BibitemOpen
  \bibfield  {author} {\bibinfo {author} {\bibfnamefont {Y.}~\bibnamefont
  {Wang}}, \bibinfo {author} {\bibfnamefont {J.}~\bibnamefont {Lv}}, \bibinfo
  {author} {\bibfnamefont {L.}~\bibnamefont {Zhu}},\ and\ \bibinfo {author}
  {\bibfnamefont {Y.}~\bibnamefont {Ma}},\ }\href
  {https://doi.org/https://doi.org/10.1016/j.cpc.2012.05.008} {\bibfield
  {journal} {\bibinfo  {journal} {Computer Physics Communications}\ }\textbf
  {\bibinfo {volume} {183}},\ \bibinfo {pages} {2063} (\bibinfo {year}
  {2012})}\BibitemShut {NoStop}%
\bibitem [{\citenamefont {Wang}\ \emph {et~al.}(2016)\citenamefont {Wang},
  \citenamefont {Wang}, \citenamefont {Lv}, \citenamefont {Li}, \citenamefont
  {Zhang},\ and\ \citenamefont {Ma}}]{WANG2016406}%
  \BibitemOpen
  \bibfield  {author} {\bibinfo {author} {\bibfnamefont {H.}~\bibnamefont
  {Wang}}, \bibinfo {author} {\bibfnamefont {Y.}~\bibnamefont {Wang}}, \bibinfo
  {author} {\bibfnamefont {J.}~\bibnamefont {Lv}}, \bibinfo {author}
  {\bibfnamefont {Q.}~\bibnamefont {Li}}, \bibinfo {author} {\bibfnamefont
  {L.}~\bibnamefont {Zhang}},\ and\ \bibinfo {author} {\bibfnamefont
  {Y.}~\bibnamefont {Ma}},\ }\href
  {https://doi.org/https://doi.org/10.1016/j.commatsci.2015.09.037} {\bibfield
  {journal} {\bibinfo  {journal} {Computational Materials Science}\ }\textbf
  {\bibinfo {volume} {112}},\ \bibinfo {pages} {406 } (\bibinfo {year}
  {2016})}\BibitemShut {NoStop}%
\bibitem [{\citenamefont {Pickard}\ and\ \citenamefont
  {Needs}(2006)}]{PhysRevLett.97.045504}%
  \BibitemOpen
  \bibfield  {author} {\bibinfo {author} {\bibfnamefont {C.~J.}\ \bibnamefont
  {Pickard}}\ and\ \bibinfo {author} {\bibfnamefont {R.~J.}\ \bibnamefont
  {Needs}},\ }\href {https://doi.org/10.1103/PhysRevLett.97.045504} {\bibfield
  {journal} {\bibinfo  {journal} {Phys. Rev. Lett.}\ }\textbf {\bibinfo
  {volume} {97}},\ \bibinfo {pages} {045504} (\bibinfo {year}
  {2006})}\BibitemShut {NoStop}%
\bibitem [{\citenamefont {Pickard}\ and\ \citenamefont
  {Needs}(2011)}]{Pickard_2011}%
  \BibitemOpen
  \bibfield  {author} {\bibinfo {author} {\bibfnamefont {C.~J.}\ \bibnamefont
  {Pickard}}\ and\ \bibinfo {author} {\bibfnamefont {R.~J.}\ \bibnamefont
  {Needs}},\ }\href {https://doi.org/10.1088/0953-8984/23/5/053201} {\bibfield
  {journal} {\bibinfo  {journal} {Journal of Physics: Condensed Matter}\
  }\textbf {\bibinfo {volume} {23}},\ \bibinfo {pages} {053201} (\bibinfo
  {year} {2011})}\BibitemShut {NoStop}%
\bibitem [{\citenamefont {Peng}\ \emph {et~al.}(2017)\citenamefont {Peng},
  \citenamefont {Sun}, \citenamefont {Pickard}, \citenamefont {Needs},
  \citenamefont {Wu},\ and\ \citenamefont {Ma}}]{PhysRevLett.119.107001}%
  \BibitemOpen
  \bibfield  {author} {\bibinfo {author} {\bibfnamefont {F.}~\bibnamefont
  {Peng}}, \bibinfo {author} {\bibfnamefont {Y.}~\bibnamefont {Sun}}, \bibinfo
  {author} {\bibfnamefont {C.~J.}\ \bibnamefont {Pickard}}, \bibinfo {author}
  {\bibfnamefont {R.~J.}\ \bibnamefont {Needs}}, \bibinfo {author}
  {\bibfnamefont {Q.}~\bibnamefont {Wu}},\ and\ \bibinfo {author}
  {\bibfnamefont {Y.}~\bibnamefont {Ma}},\ }\href
  {https://doi.org/10.1103/PhysRevLett.119.107001} {\bibfield  {journal}
  {\bibinfo  {journal} {Phys. Rev. Lett.}\ }\textbf {\bibinfo {volume} {119}},\
  \bibinfo {pages} {107001} (\bibinfo {year} {2017})}\BibitemShut {NoStop}%
\bibitem [{\citenamefont {Liu}\ \emph {et~al.}(2017)\citenamefont {Liu},
  \citenamefont {Naumov}, \citenamefont {Hoffmann}, \citenamefont {Ashcroft},\
  and\ \citenamefont {Hemley}}]{Liu6990}%
  \BibitemOpen
  \bibfield  {author} {\bibinfo {author} {\bibfnamefont {H.}~\bibnamefont
  {Liu}}, \bibinfo {author} {\bibfnamefont {I.~I.}\ \bibnamefont {Naumov}},
  \bibinfo {author} {\bibfnamefont {R.}~\bibnamefont {Hoffmann}}, \bibinfo
  {author} {\bibfnamefont {N.~W.}\ \bibnamefont {Ashcroft}},\ and\ \bibinfo
  {author} {\bibfnamefont {R.~J.}\ \bibnamefont {Hemley}},\ }\href
  {https://doi.org/10.1073/pnas.1704505114} {\bibfield  {journal} {\bibinfo
  {journal} {Proceedings of the National Academy of Sciences}\ }\textbf
  {\bibinfo {volume} {114}},\ \bibinfo {pages} {6990} (\bibinfo {year}
  {2017})}\BibitemShut {NoStop}%
\bibitem [{\citenamefont {Geballe}\ \emph {et~al.}(2018)\citenamefont
  {Geballe}, \citenamefont {Liu}, \citenamefont {Mishra}, \citenamefont
  {Ahart}, \citenamefont {Somayazulu}, \citenamefont {Meng}, \citenamefont
  {Baldini},\ and\ \citenamefont {Hemley}}]{doi:10.1002/anie.201709970}%
  \BibitemOpen
  \bibfield  {author} {\bibinfo {author} {\bibfnamefont {Z.~M.}\ \bibnamefont
  {Geballe}}, \bibinfo {author} {\bibfnamefont {H.}~\bibnamefont {Liu}},
  \bibinfo {author} {\bibfnamefont {A.~K.}\ \bibnamefont {Mishra}}, \bibinfo
  {author} {\bibfnamefont {M.}~\bibnamefont {Ahart}}, \bibinfo {author}
  {\bibfnamefont {M.}~\bibnamefont {Somayazulu}}, \bibinfo {author}
  {\bibfnamefont {Y.}~\bibnamefont {Meng}}, \bibinfo {author} {\bibfnamefont
  {M.}~\bibnamefont {Baldini}},\ and\ \bibinfo {author} {\bibfnamefont {R.~J.}\
  \bibnamefont {Hemley}},\ }\href {https://doi.org/10.1002/anie.201709970}
  {\bibfield  {journal} {\bibinfo  {journal} {Angewandte Chemie International
  Edition}\ }\textbf {\bibinfo {volume} {57}},\ \bibinfo {pages} {688}
  (\bibinfo {year} {2018})}\BibitemShut {NoStop}%
\bibitem [{\citenamefont {Drozdov}\ \emph {et~al.}(2019)\citenamefont
  {Drozdov}, \citenamefont {Kong}, \citenamefont {Minkov}, \citenamefont
  {Besedin}, \citenamefont {Kuzovnikov}, \citenamefont {Mozaffari},
  \citenamefont {Balicas}, \citenamefont {Balakirev}, \citenamefont {Graf},
  \citenamefont {Prakapenka}, \citenamefont {Greenberg}, \citenamefont
  {Knyazev}, \citenamefont {Tkacz},\ and\ \citenamefont
  {Eremets}}]{Drozdov2019}%
  \BibitemOpen
  \bibfield  {author} {\bibinfo {author} {\bibfnamefont {A.~P.}\ \bibnamefont
  {Drozdov}}, \bibinfo {author} {\bibfnamefont {P.~P.}\ \bibnamefont {Kong}},
  \bibinfo {author} {\bibfnamefont {V.~S.}\ \bibnamefont {Minkov}}, \bibinfo
  {author} {\bibfnamefont {S.~P.}\ \bibnamefont {Besedin}}, \bibinfo {author}
  {\bibfnamefont {M.~A.}\ \bibnamefont {Kuzovnikov}}, \bibinfo {author}
  {\bibfnamefont {S.}~\bibnamefont {Mozaffari}}, \bibinfo {author}
  {\bibfnamefont {L.}~\bibnamefont {Balicas}}, \bibinfo {author} {\bibfnamefont
  {F.~F.}\ \bibnamefont {Balakirev}}, \bibinfo {author} {\bibfnamefont {D.~E.}\
  \bibnamefont {Graf}}, \bibinfo {author} {\bibfnamefont {V.~B.}\ \bibnamefont
  {Prakapenka}}, \bibinfo {author} {\bibfnamefont {E.}~\bibnamefont
  {Greenberg}}, \bibinfo {author} {\bibfnamefont {D.~A.}\ \bibnamefont
  {Knyazev}}, \bibinfo {author} {\bibfnamefont {M.}~\bibnamefont {Tkacz}},\
  and\ \bibinfo {author} {\bibfnamefont {M.~I.}\ \bibnamefont {Eremets}},\
  }\href {https://doi.org/10.1038/s41586-019-1201-8} {\bibfield  {journal}
  {\bibinfo  {journal} {Nature}\ }\textbf {\bibinfo {volume} {569}},\ \bibinfo
  {pages} {528} (\bibinfo {year} {2019})}\BibitemShut {NoStop}%
\bibitem [{\citenamefont {Somayazulu}\ \emph {et~al.}(2019)\citenamefont
  {Somayazulu}, \citenamefont {Ahart}, \citenamefont {Mishra}, \citenamefont
  {Geballe}, \citenamefont {Baldini}, \citenamefont {Meng}, \citenamefont
  {Struzhkin},\ and\ \citenamefont {Hemley}}]{PhysRevLett.122.027001}%
  \BibitemOpen
  \bibfield  {author} {\bibinfo {author} {\bibfnamefont {M.}~\bibnamefont
  {Somayazulu}}, \bibinfo {author} {\bibfnamefont {M.}~\bibnamefont {Ahart}},
  \bibinfo {author} {\bibfnamefont {A.~K.}\ \bibnamefont {Mishra}}, \bibinfo
  {author} {\bibfnamefont {Z.~M.}\ \bibnamefont {Geballe}}, \bibinfo {author}
  {\bibfnamefont {M.}~\bibnamefont {Baldini}}, \bibinfo {author} {\bibfnamefont
  {Y.}~\bibnamefont {Meng}}, \bibinfo {author} {\bibfnamefont {V.~V.}\
  \bibnamefont {Struzhkin}},\ and\ \bibinfo {author} {\bibfnamefont {R.~J.}\
  \bibnamefont {Hemley}},\ }\href
  {https://doi.org/10.1103/PhysRevLett.122.027001} {\bibfield  {journal}
  {\bibinfo  {journal} {Phys. Rev. Lett.}\ }\textbf {\bibinfo {volume} {122}},\
  \bibinfo {pages} {027001} (\bibinfo {year} {2019})}\BibitemShut {NoStop}%
\bibitem [{\citenamefont {Struzhkin}\ \emph {et~al.}(2020)\citenamefont
  {Struzhkin}, \citenamefont {Li}, \citenamefont {Ji}, \citenamefont {Chen},
  \citenamefont {Prakapenka}, \citenamefont {Greenberg}, \citenamefont
  {Troyan}, \citenamefont {Gavriliuk},\ and\ \citenamefont
  {Mao}}]{doi:10.1063/1.5128736}%
  \BibitemOpen
  \bibfield  {author} {\bibinfo {author} {\bibfnamefont {V.}~\bibnamefont
  {Struzhkin}}, \bibinfo {author} {\bibfnamefont {B.}~\bibnamefont {Li}},
  \bibinfo {author} {\bibfnamefont {C.}~\bibnamefont {Ji}}, \bibinfo {author}
  {\bibfnamefont {X.-J.}\ \bibnamefont {Chen}}, \bibinfo {author}
  {\bibfnamefont {V.}~\bibnamefont {Prakapenka}}, \bibinfo {author}
  {\bibfnamefont {E.}~\bibnamefont {Greenberg}}, \bibinfo {author}
  {\bibfnamefont {I.}~\bibnamefont {Troyan}}, \bibinfo {author} {\bibfnamefont
  {A.}~\bibnamefont {Gavriliuk}},\ and\ \bibinfo {author} {\bibfnamefont
  {H.-k.}\ \bibnamefont {Mao}},\ }\href {https://doi.org/10.1063/1.5128736}
  {\bibfield  {journal} {\bibinfo  {journal} {Matter and Radiation at
  Extremes}\ }\textbf {\bibinfo {volume} {5}},\ \bibinfo {pages} {028201}
  (\bibinfo {year} {2020})}\BibitemShut {NoStop}%
\bibitem [{\citenamefont {Troyan}\ \emph {et~al.}(2021)\citenamefont {Troyan},
  \citenamefont {Semenok}, \citenamefont {Kvashnin}, \citenamefont {Sadakov},
  \citenamefont {Sobolevskiy}, \citenamefont {Pudalov}, \citenamefont
  {Ivanova}, \citenamefont {Prakapenka}, \citenamefont {Greenberg},
  \citenamefont {Gavriliuk}, \citenamefont {Lyubutin}, \citenamefont
  {Struzhkin}, \citenamefont {Bergara}, \citenamefont {Errea}, \citenamefont
  {Bianco}, \citenamefont {Calandra}, \citenamefont {Mauri}, \citenamefont
  {Monacelli}, \citenamefont {Akashi},\ and\ \citenamefont
  {Oganov}}]{https://doi.org/10.1002/adma.202006832}%
  \BibitemOpen
  \bibfield  {author} {\bibinfo {author} {\bibfnamefont {I.~A.}\ \bibnamefont
  {Troyan}}, \bibinfo {author} {\bibfnamefont {D.~V.}\ \bibnamefont {Semenok}},
  \bibinfo {author} {\bibfnamefont {A.~G.}\ \bibnamefont {Kvashnin}}, \bibinfo
  {author} {\bibfnamefont {A.~V.}\ \bibnamefont {Sadakov}}, \bibinfo {author}
  {\bibfnamefont {O.~A.}\ \bibnamefont {Sobolevskiy}}, \bibinfo {author}
  {\bibfnamefont {V.~M.}\ \bibnamefont {Pudalov}}, \bibinfo {author}
  {\bibfnamefont {A.~G.}\ \bibnamefont {Ivanova}}, \bibinfo {author}
  {\bibfnamefont {V.~B.}\ \bibnamefont {Prakapenka}}, \bibinfo {author}
  {\bibfnamefont {E.}~\bibnamefont {Greenberg}}, \bibinfo {author}
  {\bibfnamefont {A.~G.}\ \bibnamefont {Gavriliuk}}, \bibinfo {author}
  {\bibfnamefont {I.~S.}\ \bibnamefont {Lyubutin}}, \bibinfo {author}
  {\bibfnamefont {V.~V.}\ \bibnamefont {Struzhkin}}, \bibinfo {author}
  {\bibfnamefont {A.}~\bibnamefont {Bergara}}, \bibinfo {author} {\bibfnamefont
  {I.}~\bibnamefont {Errea}}, \bibinfo {author} {\bibfnamefont
  {R.}~\bibnamefont {Bianco}}, \bibinfo {author} {\bibfnamefont
  {M.}~\bibnamefont {Calandra}}, \bibinfo {author} {\bibfnamefont
  {F.}~\bibnamefont {Mauri}}, \bibinfo {author} {\bibfnamefont
  {L.}~\bibnamefont {Monacelli}}, \bibinfo {author} {\bibfnamefont
  {R.}~\bibnamefont {Akashi}},\ and\ \bibinfo {author} {\bibfnamefont {A.~R.}\
  \bibnamefont {Oganov}},\ }\href
  {https://doi.org/https://doi.org/10.1002/adma.202006832} {\bibfield
  {journal} {\bibinfo  {journal} {Advanced Materials}\ }\textbf {\bibinfo
  {volume} {33}},\ \bibinfo {pages} {2006832} (\bibinfo {year}
  {2021})}\BibitemShut {NoStop}%
\bibitem [{\citenamefont {Sun}\ \emph {et~al.}(2019)\citenamefont {Sun},
  \citenamefont {Lv}, \citenamefont {Xie}, \citenamefont {Liu},\ and\
  \citenamefont {Ma}}]{PhysRevLett.123.097001}%
  \BibitemOpen
  \bibfield  {author} {\bibinfo {author} {\bibfnamefont {Y.}~\bibnamefont
  {Sun}}, \bibinfo {author} {\bibfnamefont {J.}~\bibnamefont {Lv}}, \bibinfo
  {author} {\bibfnamefont {Y.}~\bibnamefont {Xie}}, \bibinfo {author}
  {\bibfnamefont {H.}~\bibnamefont {Liu}},\ and\ \bibinfo {author}
  {\bibfnamefont {Y.}~\bibnamefont {Ma}},\ }\href
  {https://doi.org/10.1103/PhysRevLett.123.097001} {\bibfield  {journal}
  {\bibinfo  {journal} {Phys. Rev. Lett.}\ }\textbf {\bibinfo {volume} {123}},\
  \bibinfo {pages} {097001} (\bibinfo {year} {2019})}\BibitemShut {NoStop}%
\bibitem [{\citenamefont {Ge}\ \emph {et~al.}(2016)\citenamefont {Ge},
  \citenamefont {Zhang},\ and\ \citenamefont {Yao}}]{PhysRevB.93.224513}%
  \BibitemOpen
  \bibfield  {author} {\bibinfo {author} {\bibfnamefont {Y.}~\bibnamefont
  {Ge}}, \bibinfo {author} {\bibfnamefont {F.}~\bibnamefont {Zhang}},\ and\
  \bibinfo {author} {\bibfnamefont {Y.}~\bibnamefont {Yao}},\ }\href
  {https://doi.org/10.1103/PhysRevB.93.224513} {\bibfield  {journal} {\bibinfo
  {journal} {Phys. Rev. B}\ }\textbf {\bibinfo {volume} {93}},\ \bibinfo
  {pages} {224513} (\bibinfo {year} {2016})}\BibitemShut {NoStop}%
\bibitem [{\citenamefont {Zhang}\ \emph {et~al.}(2016)\citenamefont {Zhang},
  \citenamefont {Zhu}, \citenamefont {Liu},\ and\ \citenamefont
  {Yang}}]{doi:10.1021/acs.inorgchem.6b01949}%
  \BibitemOpen
  \bibfield  {author} {\bibinfo {author} {\bibfnamefont {S.}~\bibnamefont
  {Zhang}}, \bibinfo {author} {\bibfnamefont {L.}~\bibnamefont {Zhu}}, \bibinfo
  {author} {\bibfnamefont {H.}~\bibnamefont {Liu}},\ and\ \bibinfo {author}
  {\bibfnamefont {G.}~\bibnamefont {Yang}},\ }\href
  {https://doi.org/10.1021/acs.inorgchem.6b01949} {\bibfield  {journal}
  {\bibinfo  {journal} {Inorganic Chemistry}\ }\textbf {\bibinfo {volume}
  {55}},\ \bibinfo {pages} {11434} (\bibinfo {year} {2016})}\BibitemShut
  {NoStop}%
\bibitem [{\citenamefont {Ma}\ \emph {et~al.}(2017{\natexlab{a}})\citenamefont
  {Ma}, \citenamefont {Duan}, \citenamefont {Shao}, \citenamefont {Yu},
  \citenamefont {Liu}, \citenamefont {Tian}, \citenamefont {Huang},
  \citenamefont {Li}, \citenamefont {Liu},\ and\ \citenamefont
  {Cui}}]{PhysRevB.96.144518}%
  \BibitemOpen
  \bibfield  {author} {\bibinfo {author} {\bibfnamefont {Y.}~\bibnamefont
  {Ma}}, \bibinfo {author} {\bibfnamefont {D.}~\bibnamefont {Duan}}, \bibinfo
  {author} {\bibfnamefont {Z.}~\bibnamefont {Shao}}, \bibinfo {author}
  {\bibfnamefont {H.}~\bibnamefont {Yu}}, \bibinfo {author} {\bibfnamefont
  {H.}~\bibnamefont {Liu}}, \bibinfo {author} {\bibfnamefont {F.}~\bibnamefont
  {Tian}}, \bibinfo {author} {\bibfnamefont {X.}~\bibnamefont {Huang}},
  \bibinfo {author} {\bibfnamefont {D.}~\bibnamefont {Li}}, \bibinfo {author}
  {\bibfnamefont {B.}~\bibnamefont {Liu}},\ and\ \bibinfo {author}
  {\bibfnamefont {T.}~\bibnamefont {Cui}},\ }\href
  {https://doi.org/10.1103/PhysRevB.96.144518} {\bibfield  {journal} {\bibinfo
  {journal} {Phys. Rev. B}\ }\textbf {\bibinfo {volume} {96}},\ \bibinfo
  {pages} {144518} (\bibinfo {year} {2017}{\natexlab{a}})}\BibitemShut
  {NoStop}%
\bibitem [{\citenamefont {Ma}\ \emph {et~al.}(2017{\natexlab{b}})\citenamefont
  {Ma}, \citenamefont {Duan}, \citenamefont {Shao}, \citenamefont {Li},
  \citenamefont {Wang}, \citenamefont {Yu}, \citenamefont {Tian}, \citenamefont
  {Xie}, \citenamefont {Liu},\ and\ \citenamefont {Cui}}]{C7CP05267G}%
  \BibitemOpen
  \bibfield  {author} {\bibinfo {author} {\bibfnamefont {Y.}~\bibnamefont
  {Ma}}, \bibinfo {author} {\bibfnamefont {D.}~\bibnamefont {Duan}}, \bibinfo
  {author} {\bibfnamefont {Z.}~\bibnamefont {Shao}}, \bibinfo {author}
  {\bibfnamefont {D.}~\bibnamefont {Li}}, \bibinfo {author} {\bibfnamefont
  {L.}~\bibnamefont {Wang}}, \bibinfo {author} {\bibfnamefont {H.}~\bibnamefont
  {Yu}}, \bibinfo {author} {\bibfnamefont {F.}~\bibnamefont {Tian}}, \bibinfo
  {author} {\bibfnamefont {H.}~\bibnamefont {Xie}}, \bibinfo {author}
  {\bibfnamefont {B.}~\bibnamefont {Liu}},\ and\ \bibinfo {author}
  {\bibfnamefont {T.}~\bibnamefont {Cui}},\ }\href
  {https://doi.org/10.1039/C7CP05267G} {\bibfield  {journal} {\bibinfo
  {journal} {Phys. Chem. Chem. Phys.}\ }\textbf {\bibinfo {volume} {19}},\
  \bibinfo {pages} {27406} (\bibinfo {year} {2017}{\natexlab{b}})}\BibitemShut
  {NoStop}%
\bibitem [{\citenamefont {Rahm}\ \emph {et~al.}(2017)\citenamefont {Rahm},
  \citenamefont {Hoffmann},\ and\ \citenamefont
  {Ashcroft}}]{doi:10.1021/jacs.7b04456}%
  \BibitemOpen
  \bibfield  {author} {\bibinfo {author} {\bibfnamefont {M.}~\bibnamefont
  {Rahm}}, \bibinfo {author} {\bibfnamefont {R.}~\bibnamefont {Hoffmann}},\
  and\ \bibinfo {author} {\bibfnamefont {N.~W.}\ \bibnamefont {Ashcroft}},\
  }\href {https://doi.org/10.1021/jacs.7b04456} {\bibfield  {journal} {\bibinfo
   {journal} {Journal of the American Chemical Society}\ }\textbf {\bibinfo
  {volume} {139}},\ \bibinfo {pages} {8740} (\bibinfo {year}
  {2017})}\BibitemShut {NoStop}%
\bibitem [{\citenamefont {Liu}\ \emph {et~al.}(2018)\citenamefont {Liu},
  \citenamefont {Cui}, \citenamefont {Shi}, \citenamefont {Zhu}, \citenamefont
  {Chen}, \citenamefont {Lin}, \citenamefont {Su}, \citenamefont {Ma},
  \citenamefont {Yang}, \citenamefont {Xu}, \citenamefont {Hao}, \citenamefont
  {Durajski}, \citenamefont {Qi}, \citenamefont {Li},\ and\ \citenamefont
  {Li}}]{PhysRevB.98.174101}%
  \BibitemOpen
  \bibfield  {author} {\bibinfo {author} {\bibfnamefont {B.}~\bibnamefont
  {Liu}}, \bibinfo {author} {\bibfnamefont {W.}~\bibnamefont {Cui}}, \bibinfo
  {author} {\bibfnamefont {J.}~\bibnamefont {Shi}}, \bibinfo {author}
  {\bibfnamefont {L.}~\bibnamefont {Zhu}}, \bibinfo {author} {\bibfnamefont
  {J.}~\bibnamefont {Chen}}, \bibinfo {author} {\bibfnamefont {S.}~\bibnamefont
  {Lin}}, \bibinfo {author} {\bibfnamefont {R.}~\bibnamefont {Su}}, \bibinfo
  {author} {\bibfnamefont {J.}~\bibnamefont {Ma}}, \bibinfo {author}
  {\bibfnamefont {K.}~\bibnamefont {Yang}}, \bibinfo {author} {\bibfnamefont
  {M.}~\bibnamefont {Xu}}, \bibinfo {author} {\bibfnamefont {J.}~\bibnamefont
  {Hao}}, \bibinfo {author} {\bibfnamefont {A.~P.}\ \bibnamefont {Durajski}},
  \bibinfo {author} {\bibfnamefont {J.}~\bibnamefont {Qi}}, \bibinfo {author}
  {\bibfnamefont {Y.}~\bibnamefont {Li}},\ and\ \bibinfo {author}
  {\bibfnamefont {Y.}~\bibnamefont {Li}},\ }\href
  {https://doi.org/10.1103/PhysRevB.98.174101} {\bibfield  {journal} {\bibinfo
  {journal} {Phys. Rev. B}\ }\textbf {\bibinfo {volume} {98}},\ \bibinfo
  {pages} {174101} (\bibinfo {year} {2018})}\BibitemShut {NoStop}%
\bibitem [{\citenamefont {Amsler}(2019)}]{PhysRevB.99.060102}%
  \BibitemOpen
  \bibfield  {author} {\bibinfo {author} {\bibfnamefont {M.}~\bibnamefont
  {Amsler}},\ }\href {https://doi.org/10.1103/PhysRevB.99.060102} {\bibfield
  {journal} {\bibinfo  {journal} {Phys. Rev. B}\ }\textbf {\bibinfo {volume}
  {99}},\ \bibinfo {pages} {060102} (\bibinfo {year} {2019})}\BibitemShut
  {NoStop}%
\bibitem [{\citenamefont {Liang}\ \emph
  {et~al.}(2019{\natexlab{a}})\citenamefont {Liang}, \citenamefont {Bergara},
  \citenamefont {Wang}, \citenamefont {Wen}, \citenamefont {Zhao},
  \citenamefont {Zhou}, \citenamefont {He}, \citenamefont {Gao},\ and\
  \citenamefont {Tian}}]{PhysRevB.99.100505}%
  \BibitemOpen
  \bibfield  {author} {\bibinfo {author} {\bibfnamefont {X.}~\bibnamefont
  {Liang}}, \bibinfo {author} {\bibfnamefont {A.}~\bibnamefont {Bergara}},
  \bibinfo {author} {\bibfnamefont {L.}~\bibnamefont {Wang}}, \bibinfo {author}
  {\bibfnamefont {B.}~\bibnamefont {Wen}}, \bibinfo {author} {\bibfnamefont
  {Z.}~\bibnamefont {Zhao}}, \bibinfo {author} {\bibfnamefont {X.-F.}\
  \bibnamefont {Zhou}}, \bibinfo {author} {\bibfnamefont {J.}~\bibnamefont
  {He}}, \bibinfo {author} {\bibfnamefont {G.}~\bibnamefont {Gao}},\ and\
  \bibinfo {author} {\bibfnamefont {Y.}~\bibnamefont {Tian}},\ }\href
  {https://doi.org/10.1103/PhysRevB.99.100505} {\bibfield  {journal} {\bibinfo
  {journal} {Phys. Rev. B}\ }\textbf {\bibinfo {volume} {99}},\ \bibinfo
  {pages} {100505} (\bibinfo {year} {2019}{\natexlab{a}})}\BibitemShut
  {NoStop}%
\bibitem [{\citenamefont {Xie}\ \emph {et~al.}(2019)\citenamefont {Xie},
  \citenamefont {Duan}, \citenamefont {Shao}, \citenamefont {Song},
  \citenamefont {Wang}, \citenamefont {Xiao}, \citenamefont {Li}, \citenamefont
  {Tian}, \citenamefont {Liu},\ and\ \citenamefont {Cui}}]{Xie_2019}%
  \BibitemOpen
  \bibfield  {author} {\bibinfo {author} {\bibfnamefont {H.}~\bibnamefont
  {Xie}}, \bibinfo {author} {\bibfnamefont {D.}~\bibnamefont {Duan}}, \bibinfo
  {author} {\bibfnamefont {Z.}~\bibnamefont {Shao}}, \bibinfo {author}
  {\bibfnamefont {H.}~\bibnamefont {Song}}, \bibinfo {author} {\bibfnamefont
  {Y.}~\bibnamefont {Wang}}, \bibinfo {author} {\bibfnamefont {X.}~\bibnamefont
  {Xiao}}, \bibinfo {author} {\bibfnamefont {D.}~\bibnamefont {Li}}, \bibinfo
  {author} {\bibfnamefont {F.}~\bibnamefont {Tian}}, \bibinfo {author}
  {\bibfnamefont {B.}~\bibnamefont {Liu}},\ and\ \bibinfo {author}
  {\bibfnamefont {T.}~\bibnamefont {Cui}},\ }\href
  {https://doi.org/10.1088/1361-648x/ab09b4} {\bibfield  {journal} {\bibinfo
  {journal} {Journal of Physics: Condensed Matter}\ }\textbf {\bibinfo {volume}
  {31}},\ \bibinfo {pages} {245404} (\bibinfo {year} {2019})}\BibitemShut
  {NoStop}%
\bibitem [{\citenamefont {Liang}\ \emph
  {et~al.}(2019{\natexlab{b}})\citenamefont {Liang}, \citenamefont {Zhao},
  \citenamefont {Shao}, \citenamefont {Bergara}, \citenamefont {Liu},
  \citenamefont {Wang}, \citenamefont {Sun}, \citenamefont {Zhang},
  \citenamefont {Gao}, \citenamefont {Zhao}, \citenamefont {Zhou},
  \citenamefont {He}, \citenamefont {Yu}, \citenamefont {Gao},\ and\
  \citenamefont {Tian}}]{PhysRevB.100.184502}%
  \BibitemOpen
  \bibfield  {author} {\bibinfo {author} {\bibfnamefont {X.}~\bibnamefont
  {Liang}}, \bibinfo {author} {\bibfnamefont {S.}~\bibnamefont {Zhao}},
  \bibinfo {author} {\bibfnamefont {C.}~\bibnamefont {Shao}}, \bibinfo {author}
  {\bibfnamefont {A.}~\bibnamefont {Bergara}}, \bibinfo {author} {\bibfnamefont
  {H.}~\bibnamefont {Liu}}, \bibinfo {author} {\bibfnamefont {L.}~\bibnamefont
  {Wang}}, \bibinfo {author} {\bibfnamefont {R.}~\bibnamefont {Sun}}, \bibinfo
  {author} {\bibfnamefont {Y.}~\bibnamefont {Zhang}}, \bibinfo {author}
  {\bibfnamefont {Y.}~\bibnamefont {Gao}}, \bibinfo {author} {\bibfnamefont
  {Z.}~\bibnamefont {Zhao}}, \bibinfo {author} {\bibfnamefont {X.-F.}\
  \bibnamefont {Zhou}}, \bibinfo {author} {\bibfnamefont {J.}~\bibnamefont
  {He}}, \bibinfo {author} {\bibfnamefont {D.}~\bibnamefont {Yu}}, \bibinfo
  {author} {\bibfnamefont {G.}~\bibnamefont {Gao}},\ and\ \bibinfo {author}
  {\bibfnamefont {Y.}~\bibnamefont {Tian}},\ }\href
  {https://doi.org/10.1103/PhysRevB.100.184502} {\bibfield  {journal} {\bibinfo
   {journal} {Phys. Rev. B}\ }\textbf {\bibinfo {volume} {100}},\ \bibinfo
  {pages} {184502} (\bibinfo {year} {2019}{\natexlab{b}})}\BibitemShut
  {NoStop}%
\bibitem [{\citenamefont {Shao}\ \emph {et~al.}(2019)\citenamefont {Shao},
  \citenamefont {Duan}, \citenamefont {Ma}, \citenamefont {Yu}, \citenamefont
  {Song}, \citenamefont {Xie}, \citenamefont {Li}, \citenamefont {Tian},
  \citenamefont {Liu},\ and\ \citenamefont {Cui}}]{10.1038/s41524-019-0244-6}%
  \BibitemOpen
  \bibfield  {author} {\bibinfo {author} {\bibfnamefont {Z.}~\bibnamefont
  {Shao}}, \bibinfo {author} {\bibfnamefont {D.}~\bibnamefont {Duan}}, \bibinfo
  {author} {\bibfnamefont {Y.}~\bibnamefont {Ma}}, \bibinfo {author}
  {\bibfnamefont {H.}~\bibnamefont {Yu}}, \bibinfo {author} {\bibfnamefont
  {H.}~\bibnamefont {Song}}, \bibinfo {author} {\bibfnamefont {H.}~\bibnamefont
  {Xie}}, \bibinfo {author} {\bibfnamefont {D.}~\bibnamefont {Li}}, \bibinfo
  {author} {\bibfnamefont {F.}~\bibnamefont {Tian}}, \bibinfo {author}
  {\bibfnamefont {B.}~\bibnamefont {Liu}},\ and\ \bibinfo {author}
  {\bibfnamefont {T.}~\bibnamefont {Cui}},\ }\href
  {https://doi.org/10.1038/s41524-019-0244-6} {\bibfield  {journal} {\bibinfo
  {journal} {npj Computational Materials}\ }\textbf {\bibinfo {volume} {5}},\
  \bibinfo {pages} {104} (\bibinfo {year} {2019})}\BibitemShut {NoStop}%
\bibitem [{\citenamefont {Zhang}\ \emph {et~al.}(2020)\citenamefont {Zhang},
  \citenamefont {Sun}, \citenamefont {Li}, \citenamefont {Lv},\ and\
  \citenamefont {Liu}}]{PhysRevB.102.184103}%
  \BibitemOpen
  \bibfield  {author} {\bibinfo {author} {\bibfnamefont {P.}~\bibnamefont
  {Zhang}}, \bibinfo {author} {\bibfnamefont {Y.}~\bibnamefont {Sun}}, \bibinfo
  {author} {\bibfnamefont {X.}~\bibnamefont {Li}}, \bibinfo {author}
  {\bibfnamefont {J.}~\bibnamefont {Lv}},\ and\ \bibinfo {author}
  {\bibfnamefont {H.}~\bibnamefont {Liu}},\ }\href
  {https://doi.org/10.1103/PhysRevB.102.184103} {\bibfield  {journal} {\bibinfo
   {journal} {Phys. Rev. B}\ }\textbf {\bibinfo {volume} {102}},\ \bibinfo
  {pages} {184103} (\bibinfo {year} {2020})}\BibitemShut {NoStop}%
\bibitem [{\citenamefont {Di~Cataldo}\ \emph {et~al.}(2020)\citenamefont
  {Di~Cataldo}, \citenamefont {von~der Linden},\ and\ \citenamefont
  {Boeri}}]{PhysRevB.102.014516}%
  \BibitemOpen
  \bibfield  {author} {\bibinfo {author} {\bibfnamefont {S.}~\bibnamefont
  {Di~Cataldo}}, \bibinfo {author} {\bibfnamefont {W.}~\bibnamefont {von~der
  Linden}},\ and\ \bibinfo {author} {\bibfnamefont {L.}~\bibnamefont {Boeri}},\
  }\href {https://doi.org/10.1103/PhysRevB.102.014516} {\bibfield  {journal}
  {\bibinfo  {journal} {Phys. Rev. B}\ }\textbf {\bibinfo {volume} {102}},\
  \bibinfo {pages} {014516} (\bibinfo {year} {2020})}\BibitemShut {NoStop}%
\bibitem [{\citenamefont {Cui}\ \emph {et~al.}(2020)\citenamefont {Cui},
  \citenamefont {Bi}, \citenamefont {Shi}, \citenamefont {Li}, \citenamefont
  {Liu}, \citenamefont {Zurek},\ and\ \citenamefont
  {Hemley}}]{PhysRevB.101.134504}%
  \BibitemOpen
  \bibfield  {author} {\bibinfo {author} {\bibfnamefont {W.}~\bibnamefont
  {Cui}}, \bibinfo {author} {\bibfnamefont {T.}~\bibnamefont {Bi}}, \bibinfo
  {author} {\bibfnamefont {J.}~\bibnamefont {Shi}}, \bibinfo {author}
  {\bibfnamefont {Y.}~\bibnamefont {Li}}, \bibinfo {author} {\bibfnamefont
  {H.}~\bibnamefont {Liu}}, \bibinfo {author} {\bibfnamefont {E.}~\bibnamefont
  {Zurek}},\ and\ \bibinfo {author} {\bibfnamefont {R.~J.}\ \bibnamefont
  {Hemley}},\ }\href {https://doi.org/10.1103/PhysRevB.101.134504} {\bibfield
  {journal} {\bibinfo  {journal} {Phys. Rev. B}\ }\textbf {\bibinfo {volume}
  {101}},\ \bibinfo {pages} {134504} (\bibinfo {year} {2020})}\BibitemShut
  {NoStop}%
\bibitem [{\citenamefont {Yan}\ \emph {et~al.}(2020)\citenamefont {Yan},
  \citenamefont {Bi}, \citenamefont {Geng}, \citenamefont {Wang},\ and\
  \citenamefont {Zurek}}]{doi:10.1021/acs.jpclett.0c02299}%
  \BibitemOpen
  \bibfield  {author} {\bibinfo {author} {\bibfnamefont {Y.}~\bibnamefont
  {Yan}}, \bibinfo {author} {\bibfnamefont {T.}~\bibnamefont {Bi}}, \bibinfo
  {author} {\bibfnamefont {N.}~\bibnamefont {Geng}}, \bibinfo {author}
  {\bibfnamefont {X.}~\bibnamefont {Wang}},\ and\ \bibinfo {author}
  {\bibfnamefont {E.}~\bibnamefont {Zurek}},\ }\href
  {https://doi.org/10.1021/acs.jpclett.0c02299} {\bibfield  {journal} {\bibinfo
   {journal} {The Journal of Physical Chemistry Letters}\ }\textbf {\bibinfo
  {volume} {11}},\ \bibinfo {pages} {9629} (\bibinfo {year}
  {2020})}\BibitemShut {NoStop}%
\bibitem [{\citenamefont {Lv}\ \emph {et~al.}(2020)\citenamefont {Lv},
  \citenamefont {Zhang}, \citenamefont {Li}, \citenamefont {Hai}, \citenamefont
  {Lu}, \citenamefont {Li},\ and\ \citenamefont {Zhong}}]{C9CP06008A}%
  \BibitemOpen
  \bibfield  {author} {\bibinfo {author} {\bibfnamefont {H.-Y.}\ \bibnamefont
  {Lv}}, \bibinfo {author} {\bibfnamefont {S.-Y.}\ \bibnamefont {Zhang}},
  \bibinfo {author} {\bibfnamefont {M.-H.}\ \bibnamefont {Li}}, \bibinfo
  {author} {\bibfnamefont {Y.-L.}\ \bibnamefont {Hai}}, \bibinfo {author}
  {\bibfnamefont {N.}~\bibnamefont {Lu}}, \bibinfo {author} {\bibfnamefont
  {W.-J.}\ \bibnamefont {Li}},\ and\ \bibinfo {author} {\bibfnamefont {G.-H.}\
  \bibnamefont {Zhong}},\ }\href {https://doi.org/10.1039/C9CP06008A}
  {\bibfield  {journal} {\bibinfo  {journal} {Phys. Chem. Chem. Phys.}\
  }\textbf {\bibinfo {volume} {22}},\ \bibinfo {pages} {1069} (\bibinfo {year}
  {2020})}\BibitemShut {NoStop}%
\bibitem [{\citenamefont {Di~Cataldo}\ \emph {et~al.}(2021)\citenamefont
  {Di~Cataldo}, \citenamefont {Heil}, \citenamefont {von~der Linden},\ and\
  \citenamefont {Boeri}}]{PhysRevB.104.L020511}%
  \BibitemOpen
  \bibfield  {author} {\bibinfo {author} {\bibfnamefont {S.}~\bibnamefont
  {Di~Cataldo}}, \bibinfo {author} {\bibfnamefont {C.}~\bibnamefont {Heil}},
  \bibinfo {author} {\bibfnamefont {W.}~\bibnamefont {von~der Linden}},\ and\
  \bibinfo {author} {\bibfnamefont {L.}~\bibnamefont {Boeri}},\ }\href
  {https://doi.org/10.1103/PhysRevB.104.L020511} {\bibfield  {journal}
  {\bibinfo  {journal} {Phys. Rev. B}\ }\textbf {\bibinfo {volume} {104}},\
  \bibinfo {pages} {L020511} (\bibinfo {year} {2021})}\BibitemShut {NoStop}%
\bibitem [{\citenamefont {Wei}\ \emph {et~al.}(2021)\citenamefont {Wei},
  \citenamefont {Jia}, \citenamefont {Fang}, \citenamefont {Wang},
  \citenamefont {Qian}, \citenamefont {Yuan}, \citenamefont {Selvaraj},
  \citenamefont {Ji},\ and\ \citenamefont
  {Wei}}]{https://doi.org/10.1002/qua.26459}%
  \BibitemOpen
  \bibfield  {author} {\bibinfo {author} {\bibfnamefont {Y.~K.}\ \bibnamefont
  {Wei}}, \bibinfo {author} {\bibfnamefont {L.~Q.}\ \bibnamefont {Jia}},
  \bibinfo {author} {\bibfnamefont {Y.~Y.}\ \bibnamefont {Fang}}, \bibinfo
  {author} {\bibfnamefont {L.~J.}\ \bibnamefont {Wang}}, \bibinfo {author}
  {\bibfnamefont {Z.~X.}\ \bibnamefont {Qian}}, \bibinfo {author}
  {\bibfnamefont {J.~N.}\ \bibnamefont {Yuan}}, \bibinfo {author}
  {\bibfnamefont {G.}~\bibnamefont {Selvaraj}}, \bibinfo {author}
  {\bibfnamefont {G.~F.}\ \bibnamefont {Ji}},\ and\ \bibinfo {author}
  {\bibfnamefont {D.~Q.}\ \bibnamefont {Wei}},\ }\href
  {https://doi.org/https://doi.org/10.1002/qua.26459} {\bibfield  {journal}
  {\bibinfo  {journal} {International Journal of Quantum Chemistry}\ }\textbf
  {\bibinfo {volume} {121}},\ \bibinfo {pages} {e26459} (\bibinfo {year}
  {2021})}\BibitemShut {NoStop}%
\bibitem [{\citenamefont {Liu}\ \emph {et~al.}(2021)\citenamefont {Liu},
  \citenamefont {Cheng}, \citenamefont {Yang}, \citenamefont {Li},
  \citenamefont {Chen},\ and\ \citenamefont {Lu}}]{LIU2021127109}%
  \BibitemOpen
  \bibfield  {author} {\bibinfo {author} {\bibfnamefont {H.}~\bibnamefont
  {Liu}}, \bibinfo {author} {\bibfnamefont {R.}~\bibnamefont {Cheng}}, \bibinfo
  {author} {\bibfnamefont {K.}~\bibnamefont {Yang}}, \bibinfo {author}
  {\bibfnamefont {B.}~\bibnamefont {Li}}, \bibinfo {author} {\bibfnamefont
  {L.}~\bibnamefont {Chen}},\ and\ \bibinfo {author} {\bibfnamefont
  {W.}~\bibnamefont {Lu}},\ }\href
  {https://doi.org/https://doi.org/10.1016/j.physleta.2020.127109} {\bibfield
  {journal} {\bibinfo  {journal} {Physics Letters A}\ }\textbf {\bibinfo
  {volume} {390}},\ \bibinfo {pages} {127109} (\bibinfo {year}
  {2021})}\BibitemShut {NoStop}%
\bibitem [{\citenamefont {Song}\ \emph {et~al.}(2021)\citenamefont {Song},
  \citenamefont {Hou}, \citenamefont {Castro}, \citenamefont {Nakano},
  \citenamefont {Hongo}, \citenamefont {Takano},\ and\ \citenamefont
  {Maezono}}]{doi:10.1021/acs.chemmater.1c02371}%
  \BibitemOpen
  \bibfield  {author} {\bibinfo {author} {\bibfnamefont {P.}~\bibnamefont
  {Song}}, \bibinfo {author} {\bibfnamefont {Z.}~\bibnamefont {Hou}}, \bibinfo
  {author} {\bibfnamefont {P.~B.~d.}\ \bibnamefont {Castro}}, \bibinfo {author}
  {\bibfnamefont {K.}~\bibnamefont {Nakano}}, \bibinfo {author} {\bibfnamefont
  {K.}~\bibnamefont {Hongo}}, \bibinfo {author} {\bibfnamefont
  {Y.}~\bibnamefont {Takano}},\ and\ \bibinfo {author} {\bibfnamefont
  {R.}~\bibnamefont {Maezono}},\ }\href
  {https://doi.org/10.1021/acs.chemmater.1c02371} {\bibfield  {journal}
  {\bibinfo  {journal} {Chemistry of Materials}\ }\textbf {\bibinfo {volume}
  {33}},\ \bibinfo {pages} {9501} (\bibinfo {year} {2021})}\BibitemShut
  {NoStop}%
\bibitem [{\citenamefont {Di~Cataldo}\ \emph {et~al.}(2022)\citenamefont
  {Di~Cataldo}, \citenamefont {von~der Linden},\ and\ \citenamefont
  {Boeri}}]{10.1038/s41524-021-00691-6}%
  \BibitemOpen
  \bibfield  {author} {\bibinfo {author} {\bibfnamefont {S.}~\bibnamefont
  {Di~Cataldo}}, \bibinfo {author} {\bibfnamefont {W.}~\bibnamefont {von~der
  Linden}},\ and\ \bibinfo {author} {\bibfnamefont {L.}~\bibnamefont {Boeri}},\
  }\href {https://doi.org/10.1038/s41524-021-00691-6} {\bibfield  {journal}
  {\bibinfo  {journal} {npj Computational Materials}\ }\textbf {\bibinfo
  {volume} {8}},\ \bibinfo {pages} {2} (\bibinfo {year} {2022})}\BibitemShut
  {NoStop}%
\bibitem [{\citenamefont {Zhang}\ \emph
  {et~al.}(2022{\natexlab{a}})\citenamefont {Zhang}, \citenamefont {Cui},
  \citenamefont {Hutcheon}, \citenamefont {Shipley}, \citenamefont {Song},
  \citenamefont {Du}, \citenamefont {Kresin}, \citenamefont {Duan},
  \citenamefont {Pickard},\ and\ \citenamefont {Yao}}]{PhysRevLett.128.047001}%
  \BibitemOpen
  \bibfield  {author} {\bibinfo {author} {\bibfnamefont {Z.}~\bibnamefont
  {Zhang}}, \bibinfo {author} {\bibfnamefont {T.}~\bibnamefont {Cui}}, \bibinfo
  {author} {\bibfnamefont {M.~J.}\ \bibnamefont {Hutcheon}}, \bibinfo {author}
  {\bibfnamefont {A.~M.}\ \bibnamefont {Shipley}}, \bibinfo {author}
  {\bibfnamefont {H.}~\bibnamefont {Song}}, \bibinfo {author} {\bibfnamefont
  {M.}~\bibnamefont {Du}}, \bibinfo {author} {\bibfnamefont {V.~Z.}\
  \bibnamefont {Kresin}}, \bibinfo {author} {\bibfnamefont {D.}~\bibnamefont
  {Duan}}, \bibinfo {author} {\bibfnamefont {C.~J.}\ \bibnamefont {Pickard}},\
  and\ \bibinfo {author} {\bibfnamefont {Y.}~\bibnamefont {Yao}},\ }\href
  {https://doi.org/10.1103/PhysRevLett.128.047001} {\bibfield  {journal}
  {\bibinfo  {journal} {Phys. Rev. Lett.}\ }\textbf {\bibinfo {volume} {128}},\
  \bibinfo {pages} {047001} (\bibinfo {year} {2022}{\natexlab{a}})}\BibitemShut
  {NoStop}%
\bibitem [{\citenamefont {Zhang}\ \emph
  {et~al.}(2022{\natexlab{b}})\citenamefont {Zhang}, \citenamefont {Zhao},\
  and\ \citenamefont {Yang}}]{https://doi.org/10.1002/wcms.1582}%
  \BibitemOpen
  \bibfield  {author} {\bibinfo {author} {\bibfnamefont {X.}~\bibnamefont
  {Zhang}}, \bibinfo {author} {\bibfnamefont {Y.}~\bibnamefont {Zhao}},\ and\
  \bibinfo {author} {\bibfnamefont {G.}~\bibnamefont {Yang}},\ }\href
  {https://doi.org/https://doi.org/10.1002/wcms.1582} {\bibfield  {journal}
  {\bibinfo  {journal} {WIREs Computational Molecular Science}\ }\textbf
  {\bibinfo {volume} {12}},\ \bibinfo {pages} {e1582} (\bibinfo {year}
  {2022}{\natexlab{b}})}\BibitemShut {NoStop}%
\bibitem [{\citenamefont {Muramatsu}\ \emph {et~al.}(2015)\citenamefont
  {Muramatsu}, \citenamefont {Wanene}, \citenamefont {Somayazulu},
  \citenamefont {Vinitsky}, \citenamefont {Chandra}, \citenamefont {Strobel},
  \citenamefont {Struzhkin},\ and\ \citenamefont
  {Hemley}}]{doi:10.1021/acs.jpcc.5b03709}%
  \BibitemOpen
  \bibfield  {author} {\bibinfo {author} {\bibfnamefont {T.}~\bibnamefont
  {Muramatsu}}, \bibinfo {author} {\bibfnamefont {W.~K.}\ \bibnamefont
  {Wanene}}, \bibinfo {author} {\bibfnamefont {M.}~\bibnamefont {Somayazulu}},
  \bibinfo {author} {\bibfnamefont {E.}~\bibnamefont {Vinitsky}}, \bibinfo
  {author} {\bibfnamefont {D.}~\bibnamefont {Chandra}}, \bibinfo {author}
  {\bibfnamefont {T.~A.}\ \bibnamefont {Strobel}}, \bibinfo {author}
  {\bibfnamefont {V.~V.}\ \bibnamefont {Struzhkin}},\ and\ \bibinfo {author}
  {\bibfnamefont {R.~J.}\ \bibnamefont {Hemley}},\ }\href
  {https://doi.org/10.1021/acs.jpcc.5b03709} {\bibfield  {journal} {\bibinfo
  {journal} {The Journal of Physical Chemistry C}\ }\textbf {\bibinfo {volume}
  {119}},\ \bibinfo {pages} {18007} (\bibinfo {year} {2015})}\BibitemShut
  {NoStop}%
\bibitem [{\citenamefont {Meng}\ \emph {et~al.}(2019)\citenamefont {Meng},
  \citenamefont {Sakata}, \citenamefont {Shimizu}, \citenamefont {Iijima},
  \citenamefont {Saitoh}, \citenamefont {Sato}, \citenamefont {Takagi},\ and\
  \citenamefont {Orimo}}]{PhysRevB.99.024508}%
  \BibitemOpen
  \bibfield  {author} {\bibinfo {author} {\bibfnamefont {D.}~\bibnamefont
  {Meng}}, \bibinfo {author} {\bibfnamefont {M.}~\bibnamefont {Sakata}},
  \bibinfo {author} {\bibfnamefont {K.}~\bibnamefont {Shimizu}}, \bibinfo
  {author} {\bibfnamefont {Y.}~\bibnamefont {Iijima}}, \bibinfo {author}
  {\bibfnamefont {H.}~\bibnamefont {Saitoh}}, \bibinfo {author} {\bibfnamefont
  {T.}~\bibnamefont {Sato}}, \bibinfo {author} {\bibfnamefont {S.}~\bibnamefont
  {Takagi}},\ and\ \bibinfo {author} {\bibfnamefont {S.-i.}\ \bibnamefont
  {Orimo}},\ }\href {https://doi.org/10.1103/PhysRevB.99.024508} {\bibfield
  {journal} {\bibinfo  {journal} {Phys. Rev. B}\ }\textbf {\bibinfo {volume}
  {99}},\ \bibinfo {pages} {024508} (\bibinfo {year} {2019})}\BibitemShut
  {NoStop}%
\bibitem [{\citenamefont {Snider}\ \emph {et~al.}(2020)\citenamefont {Snider},
  \citenamefont {Dasenbrock-Gammon}, \citenamefont {McBride}, \citenamefont
  {Debessai}, \citenamefont {Vindana}, \citenamefont {Vencatasamy},
  \citenamefont {Lawler}, \citenamefont {Salamat},\ and\ \citenamefont
  {Dias}}]{10.1038/s41586-020-2801-z}%
  \BibitemOpen
  \bibfield  {author} {\bibinfo {author} {\bibfnamefont {E.}~\bibnamefont
  {Snider}}, \bibinfo {author} {\bibfnamefont {N.}~\bibnamefont
  {Dasenbrock-Gammon}}, \bibinfo {author} {\bibfnamefont {R.}~\bibnamefont
  {McBride}}, \bibinfo {author} {\bibfnamefont {M.}~\bibnamefont {Debessai}},
  \bibinfo {author} {\bibfnamefont {H.}~\bibnamefont {Vindana}}, \bibinfo
  {author} {\bibfnamefont {K.}~\bibnamefont {Vencatasamy}}, \bibinfo {author}
  {\bibfnamefont {K.~V.}\ \bibnamefont {Lawler}}, \bibinfo {author}
  {\bibfnamefont {A.}~\bibnamefont {Salamat}},\ and\ \bibinfo {author}
  {\bibfnamefont {R.~P.}\ \bibnamefont {Dias}},\ }\href
  {https://doi.org/10.1038/s41586-020-2801-z} {\bibfield  {journal} {\bibinfo
  {journal} {Nature}\ }\textbf {\bibinfo {volume} {586}},\ \bibinfo {pages}
  {373} (\bibinfo {year} {2020})}\BibitemShut {NoStop}%
\bibitem [{\citenamefont {Bardeen}\ \emph {et~al.}(1957)\citenamefont
  {Bardeen}, \citenamefont {Cooper},\ and\ \citenamefont
  {Schrieffer}}]{PhysRev.108.1175}%
  \BibitemOpen
  \bibfield  {author} {\bibinfo {author} {\bibfnamefont {J.}~\bibnamefont
  {Bardeen}}, \bibinfo {author} {\bibfnamefont {L.~N.}\ \bibnamefont
  {Cooper}},\ and\ \bibinfo {author} {\bibfnamefont {J.~R.}\ \bibnamefont
  {Schrieffer}},\ }\href {https://doi.org/10.1103/PhysRev.108.1175} {\bibfield
  {journal} {\bibinfo  {journal} {Phys. Rev.}\ }\textbf {\bibinfo {volume}
  {108}},\ \bibinfo {pages} {1175} (\bibinfo {year} {1957})}\BibitemShut
  {NoStop}%
\bibitem [{\citenamefont {Ashcroft}(2004)}]{PhysRevLett.92.187002}%
  \BibitemOpen
  \bibfield  {author} {\bibinfo {author} {\bibfnamefont {N.~W.}\ \bibnamefont
  {Ashcroft}},\ }\href {https://doi.org/10.1103/PhysRevLett.92.187002}
  {\bibfield  {journal} {\bibinfo  {journal} {Phys. Rev. Lett.}\ }\textbf
  {\bibinfo {volume} {92}},\ \bibinfo {pages} {187002} (\bibinfo {year}
  {2004})}\BibitemShut {NoStop}%
\bibitem [{\citenamefont {Ashcroft}(1968)}]{PhysRevLett.21.1748}%
  \BibitemOpen
  \bibfield  {author} {\bibinfo {author} {\bibfnamefont {N.~W.}\ \bibnamefont
  {Ashcroft}},\ }\href {https://doi.org/10.1103/PhysRevLett.21.1748} {\bibfield
   {journal} {\bibinfo  {journal} {Phys. Rev. Lett.}\ }\textbf {\bibinfo
  {volume} {21}},\ \bibinfo {pages} {1748} (\bibinfo {year}
  {1968})}\BibitemShut {NoStop}%
\bibitem [{\citenamefont {Richardson}\ and\ \citenamefont
  {Ashcroft}(1997)}]{PhysRevLett.78.118}%
  \BibitemOpen
  \bibfield  {author} {\bibinfo {author} {\bibfnamefont {C.~F.}\ \bibnamefont
  {Richardson}}\ and\ \bibinfo {author} {\bibfnamefont {N.~W.}\ \bibnamefont
  {Ashcroft}},\ }\href {https://doi.org/10.1103/PhysRevLett.78.118} {\bibfield
  {journal} {\bibinfo  {journal} {Phys. Rev. Lett.}\ }\textbf {\bibinfo
  {volume} {78}},\ \bibinfo {pages} {118} (\bibinfo {year} {1997})}\BibitemShut
  {NoStop}%
\bibitem [{\citenamefont {Dias}\ and\ \citenamefont
  {Silvera}(2017)}]{doi:10.1126/science.aal1579}%
  \BibitemOpen
  \bibfield  {author} {\bibinfo {author} {\bibfnamefont {R.~P.}\ \bibnamefont
  {Dias}}\ and\ \bibinfo {author} {\bibfnamefont {I.~F.}\ \bibnamefont
  {Silvera}},\ }\href {https://doi.org/10.1126/science.aal1579} {\bibfield
  {journal} {\bibinfo  {journal} {Science}\ }\textbf {\bibinfo {volume}
  {355}},\ \bibinfo {pages} {715} (\bibinfo {year} {2017})}\BibitemShut
  {NoStop}%
\bibitem [{\citenamefont {Koshoji}\ and\ \citenamefont
  {Ozaki}(2021{\natexlab{a}})}]{PhysRevE.104.024101}%
  \BibitemOpen
  \bibfield  {author} {\bibinfo {author} {\bibfnamefont {R.}~\bibnamefont
  {Koshoji}}\ and\ \bibinfo {author} {\bibfnamefont {T.}~\bibnamefont
  {Ozaki}},\ }\href {https://doi.org/10.1103/PhysRevE.104.024101} {\bibfield
  {journal} {\bibinfo  {journal} {Phys. Rev. E}\ }\textbf {\bibinfo {volume}
  {104}},\ \bibinfo {pages} {024101} (\bibinfo {year}
  {2021}{\natexlab{a}})}\BibitemShut {NoStop}%
\bibitem [{\citenamefont {Koshoji}\ and\ \citenamefont
  {Ozaki}(2021{\natexlab{b}})}]{koshoji2021diverse}%
  \BibitemOpen
  \bibfield  {author} {\bibinfo {author} {\bibfnamefont {R.}~\bibnamefont
  {Koshoji}}\ and\ \bibinfo {author} {\bibfnamefont {T.}~\bibnamefont
  {Ozaki}},\ }\href@noop {} {\bibinfo {title} {Diverse densest ternary sphere
  packings}} (\bibinfo {year} {2021}{\natexlab{b}}),\ \Eprint
  {https://arxiv.org/abs/2110.15505} {arXiv:2110.15505 [cond-mat.soft]}
  \BibitemShut {NoStop}%
\bibitem [{\citenamefont {Rahm}\ \emph {et~al.}(2019)\citenamefont {Rahm},
  \citenamefont {Cammi}, \citenamefont {Ashcroft},\ and\ \citenamefont
  {Hoffmann}}]{doi:10.1021/jacs.9b02634}%
  \BibitemOpen
  \bibfield  {author} {\bibinfo {author} {\bibfnamefont {M.}~\bibnamefont
  {Rahm}}, \bibinfo {author} {\bibfnamefont {R.}~\bibnamefont {Cammi}},
  \bibinfo {author} {\bibfnamefont {N.~W.}\ \bibnamefont {Ashcroft}},\ and\
  \bibinfo {author} {\bibfnamefont {R.}~\bibnamefont {Hoffmann}},\ }\href
  {https://doi.org/10.1021/jacs.9b02634} {\bibfield  {journal} {\bibinfo
  {journal} {Journal of the American Chemical Society}\ }\textbf {\bibinfo
  {volume} {141}},\ \bibinfo {pages} {10253} (\bibinfo {year}
  {2019})}\BibitemShut {NoStop}%
\bibitem [{\citenamefont {Koshoji}\ \emph {et~al.}(2021)\citenamefont
  {Koshoji}, \citenamefont {Kawamura}, \citenamefont {Fukuda},\ and\
  \citenamefont {Ozaki}}]{PhysRevE.103.023307}%
  \BibitemOpen
  \bibfield  {author} {\bibinfo {author} {\bibfnamefont {R.}~\bibnamefont
  {Koshoji}}, \bibinfo {author} {\bibfnamefont {M.}~\bibnamefont {Kawamura}},
  \bibinfo {author} {\bibfnamefont {M.}~\bibnamefont {Fukuda}},\ and\ \bibinfo
  {author} {\bibfnamefont {T.}~\bibnamefont {Ozaki}},\ }\href
  {https://doi.org/10.1103/PhysRevE.103.023307} {\bibfield  {journal} {\bibinfo
   {journal} {Phys. Rev. E}\ }\textbf {\bibinfo {volume} {103}},\ \bibinfo
  {pages} {023307} (\bibinfo {year} {2021})}\BibitemShut {NoStop}%
\bibitem [{\citenamefont {Momma}\ and\ \citenamefont
  {Izumi}(2011)}]{Momma:db5098}%
  \BibitemOpen
  \bibfield  {author} {\bibinfo {author} {\bibfnamefont {K.}~\bibnamefont
  {Momma}}\ and\ \bibinfo {author} {\bibfnamefont {F.}~\bibnamefont {Izumi}},\
  }\href@noop {} {\bibfield  {journal} {\bibinfo  {journal} {Journal of Applied
  Crystallography}\ }\textbf {\bibinfo {volume} {44}},\ \bibinfo {pages} {1272}
  (\bibinfo {year} {2011})}\BibitemShut {NoStop}%
\bibitem [{\citenamefont {Hopkins}\ \emph {et~al.}(2011)\citenamefont
  {Hopkins}, \citenamefont {Jiao}, \citenamefont {Stillinger},\ and\
  \citenamefont {Torquato}}]{PhysRevLett.107.125501}%
  \BibitemOpen
  \bibfield  {author} {\bibinfo {author} {\bibfnamefont {A.~B.}\ \bibnamefont
  {Hopkins}}, \bibinfo {author} {\bibfnamefont {Y.}~\bibnamefont {Jiao}},
  \bibinfo {author} {\bibfnamefont {F.~H.}\ \bibnamefont {Stillinger}},\ and\
  \bibinfo {author} {\bibfnamefont {S.}~\bibnamefont {Torquato}},\ }\href
  {https://doi.org/10.1103/PhysRevLett.107.125501} {\bibfield  {journal}
  {\bibinfo  {journal} {Phys. Rev. Lett.}\ }\textbf {\bibinfo {volume} {107}},\
  \bibinfo {pages} {125501} (\bibinfo {year} {2011})}\BibitemShut {NoStop}%
\bibitem [{\citenamefont {Hopkins}\ \emph {et~al.}(2012)\citenamefont
  {Hopkins}, \citenamefont {Stillinger},\ and\ \citenamefont
  {Torquato}}]{PhysRevE.85.021130}%
  \BibitemOpen
  \bibfield  {author} {\bibinfo {author} {\bibfnamefont {A.~B.}\ \bibnamefont
  {Hopkins}}, \bibinfo {author} {\bibfnamefont {F.~H.}\ \bibnamefont
  {Stillinger}},\ and\ \bibinfo {author} {\bibfnamefont {S.}~\bibnamefont
  {Torquato}},\ }\href {https://doi.org/10.1103/PhysRevE.85.021130} {\bibfield
  {journal} {\bibinfo  {journal} {Phys. Rev. E}\ }\textbf {\bibinfo {volume}
  {85}},\ \bibinfo {pages} {021130} (\bibinfo {year} {2012})}\BibitemShut
  {NoStop}%
\bibitem [{\citenamefont {Ozaki}(2003)}]{PhysRevB.67.155108}%
  \BibitemOpen
  \bibfield  {author} {\bibinfo {author} {\bibfnamefont {T.}~\bibnamefont
  {Ozaki}},\ }\href {https://doi.org/10.1103/PhysRevB.67.155108} {\bibfield
  {journal} {\bibinfo  {journal} {Phys. Rev. B}\ }\textbf {\bibinfo {volume}
  {67}},\ \bibinfo {pages} {155108} (\bibinfo {year} {2003})}\BibitemShut
  {NoStop}%
\bibitem [{\citenamefont {Ozaki}\ and\ \citenamefont
  {Kino}(2004)}]{PhysRevB.69.195113}%
  \BibitemOpen
  \bibfield  {author} {\bibinfo {author} {\bibfnamefont {T.}~\bibnamefont
  {Ozaki}}\ and\ \bibinfo {author} {\bibfnamefont {H.}~\bibnamefont {Kino}},\
  }\href {https://doi.org/10.1103/PhysRevB.69.195113} {\bibfield  {journal}
  {\bibinfo  {journal} {Phys. Rev. B}\ }\textbf {\bibinfo {volume} {69}},\
  \bibinfo {pages} {195113} (\bibinfo {year} {2004})}\BibitemShut {NoStop}%
\bibitem [{\citenamefont {Ozaki}\ and\ \citenamefont
  {Kino}(2005)}]{PhysRevB.72.045121}%
  \BibitemOpen
  \bibfield  {author} {\bibinfo {author} {\bibfnamefont {T.}~\bibnamefont
  {Ozaki}}\ and\ \bibinfo {author} {\bibfnamefont {H.}~\bibnamefont {Kino}},\
  }\href {https://doi.org/10.1103/PhysRevB.72.045121} {\bibfield  {journal}
  {\bibinfo  {journal} {Phys. Rev. B}\ }\textbf {\bibinfo {volume} {72}},\
  \bibinfo {pages} {045121} (\bibinfo {year} {2005})}\BibitemShut {NoStop}%
\bibitem [{ope()}]{openmx}%
  \BibitemOpen
  \href@noop {} {}\bibinfo {note} {T. Ozaki, H. Kino, J. Yu, M. J. Han, N.
  Kobayashi, M. Ohfuti, F. Ishii, T. Ohwaki, H. Weng, K. Terakura, 2009,
  \textbf{http://www.openmx-square.org/}}\BibitemShut {NoStop}%
\bibitem [{\citenamefont {Perdew}\ \emph {et~al.}(1996)\citenamefont {Perdew},
  \citenamefont {Burke},\ and\ \citenamefont
  {Ernzerhof}}]{PhysRevLett.77.3865}%
  \BibitemOpen
  \bibfield  {author} {\bibinfo {author} {\bibfnamefont {J.~P.}\ \bibnamefont
  {Perdew}}, \bibinfo {author} {\bibfnamefont {K.}~\bibnamefont {Burke}},\ and\
  \bibinfo {author} {\bibfnamefont {M.}~\bibnamefont {Ernzerhof}},\ }\href
  {https://doi.org/10.1103/PhysRevLett.77.3865} {\bibfield  {journal} {\bibinfo
   {journal} {Phys. Rev. Lett.}\ }\textbf {\bibinfo {volume} {77}},\ \bibinfo
  {pages} {3865} (\bibinfo {year} {1996})}\BibitemShut {NoStop}%
\bibitem [{\citenamefont {Kohn}\ and\ \citenamefont
  {Sham}(1965)}]{PhysRev.140.A1133}%
  \BibitemOpen
  \bibfield  {author} {\bibinfo {author} {\bibfnamefont {W.}~\bibnamefont
  {Kohn}}\ and\ \bibinfo {author} {\bibfnamefont {L.~J.}\ \bibnamefont
  {Sham}},\ }\href {https://doi.org/10.1103/PhysRev.140.A1133} {\bibfield
  {journal} {\bibinfo  {journal} {Phys. Rev.}\ }\textbf {\bibinfo {volume}
  {140}},\ \bibinfo {pages} {A1133} (\bibinfo {year} {1965})}\BibitemShut
  {NoStop}%
\bibitem [{sup()}]{supplemental_material}%
  \BibitemOpen
  \href@noop {} {}\bibinfo {note} {See supplemental material for details of
  pseudoatomic orbital basis funtions, phonon properties, band structures, and
  the hydrides that show the dynamical stabilities in NVTMDs.}\BibitemShut
  {Stop}%
\bibitem [{\citenamefont {Banerjee}\ \emph {et~al.}(1985)\citenamefont
  {Banerjee}, \citenamefont {Adams}, \citenamefont {Simons},\ and\
  \citenamefont {Shepard}}]{doi:10.1021/j100247a015}%
  \BibitemOpen
  \bibfield  {author} {\bibinfo {author} {\bibfnamefont {A.}~\bibnamefont
  {Banerjee}}, \bibinfo {author} {\bibfnamefont {N.}~\bibnamefont {Adams}},
  \bibinfo {author} {\bibfnamefont {J.}~\bibnamefont {Simons}},\ and\ \bibinfo
  {author} {\bibfnamefont {R.}~\bibnamefont {Shepard}},\ }\href
  {https://doi.org/10.1021/j100247a015} {\bibfield  {journal} {\bibinfo
  {journal} {The Journal of Physical Chemistry}\ }\textbf {\bibinfo {volume}
  {89}},\ \bibinfo {pages} {52} (\bibinfo {year} {1985})}\BibitemShut {NoStop}%
\bibitem [{\citenamefont {Nos\'e}(1984{\natexlab{a}})}]{doi:10.1063/1.447334}%
  \BibitemOpen
  \bibfield  {author} {\bibinfo {author} {\bibfnamefont {S.}~\bibnamefont
  {Nos\'e}},\ }\href {https://doi.org/10.1063/1.447334} {\bibfield  {journal}
  {\bibinfo  {journal} {The Journal of Chemical Physics}\ }\textbf {\bibinfo
  {volume} {81}},\ \bibinfo {pages} {511} (\bibinfo {year}
  {1984}{\natexlab{a}})}\BibitemShut {NoStop}%
\bibitem [{\citenamefont
  {Nos\'e}(1984{\natexlab{b}})}]{doi:10.1080/00268978400101201}%
  \BibitemOpen
  \bibfield  {author} {\bibinfo {author} {\bibfnamefont {S.}~\bibnamefont
  {Nos\'e}},\ }\href {https://doi.org/10.1080/00268978400101201} {\bibfield
  {journal} {\bibinfo  {journal} {Molecular Physics}\ }\textbf {\bibinfo
  {volume} {52}},\ \bibinfo {pages} {255} (\bibinfo {year}
  {1984}{\natexlab{b}})}\BibitemShut {NoStop}%
\bibitem [{\citenamefont {Hoover}(1985)}]{PhysRevA.31.1695}%
  \BibitemOpen
  \bibfield  {author} {\bibinfo {author} {\bibfnamefont {W.~G.}\ \bibnamefont
  {Hoover}},\ }\href {https://doi.org/10.1103/PhysRevA.31.1695} {\bibfield
  {journal} {\bibinfo  {journal} {Phys. Rev. A}\ }\textbf {\bibinfo {volume}
  {31}},\ \bibinfo {pages} {1695} (\bibinfo {year} {1985})}\BibitemShut
  {NoStop}%
\bibitem [{\citenamefont {Woodcock}(1971)}]{WOODCOCK1971257}%
  \BibitemOpen
  \bibfield  {author} {\bibinfo {author} {\bibfnamefont {L.}~\bibnamefont
  {Woodcock}},\ }\href
  {https://doi.org/https://doi.org/10.1016/0009-2614(71)80281-6} {\bibfield
  {journal} {\bibinfo  {journal} {Chemical Physics Letters}\ }\textbf {\bibinfo
  {volume} {10}},\ \bibinfo {pages} {257} (\bibinfo {year} {1971})}\BibitemShut
  {NoStop}%
\bibitem [{\citenamefont {Parrinello}\ and\ \citenamefont
  {Rahman}(1980)}]{PhysRevLett.45.1196}%
  \BibitemOpen
  \bibfield  {author} {\bibinfo {author} {\bibfnamefont {M.}~\bibnamefont
  {Parrinello}}\ and\ \bibinfo {author} {\bibfnamefont {A.}~\bibnamefont
  {Rahman}},\ }\href {https://doi.org/10.1103/PhysRevLett.45.1196} {\bibfield
  {journal} {\bibinfo  {journal} {Phys. Rev. Lett.}\ }\textbf {\bibinfo
  {volume} {45}},\ \bibinfo {pages} {1196} (\bibinfo {year}
  {1980})}\BibitemShut {NoStop}%
\bibitem [{\citenamefont {Tadano}\ \emph {et~al.}(2014)\citenamefont {Tadano},
  \citenamefont {Gohda},\ and\ \citenamefont {Tsuneyuki}}]{Tadano_2014}%
  \BibitemOpen
  \bibfield  {author} {\bibinfo {author} {\bibfnamefont {T.}~\bibnamefont
  {Tadano}}, \bibinfo {author} {\bibfnamefont {Y.}~\bibnamefont {Gohda}},\ and\
  \bibinfo {author} {\bibfnamefont {S.}~\bibnamefont {Tsuneyuki}},\ }\href
  {https://doi.org/10.1088/0953-8984/26/22/225402} {\bibfield  {journal}
  {\bibinfo  {journal} {Journal of Physics: Condensed Matter}\ }\textbf
  {\bibinfo {volume} {26}},\ \bibinfo {pages} {225402} (\bibinfo {year}
  {2014})}\BibitemShut {NoStop}%
\bibitem [{\citenamefont {Togo}\ and\ \citenamefont
  {Tanaka}(2018)}]{togo2018textttspglib}%
  \BibitemOpen
  \bibfield  {author} {\bibinfo {author} {\bibfnamefont {A.}~\bibnamefont
  {Togo}}\ and\ \bibinfo {author} {\bibfnamefont {I.}~\bibnamefont {Tanaka}},\
  }\href@noop {} {\bibfield  {journal} {\bibinfo  {journal} {arXiv preprint
  arXiv:1808.01590}\ } (\bibinfo {year} {2018})}\BibitemShut {NoStop}%
\bibitem [{\citenamefont {Cordero}\ \emph {et~al.}(2008)\citenamefont
  {Cordero}, \citenamefont {G\'omez}, \citenamefont {Platero-Prats},
  \citenamefont {Rev\'es}, \citenamefont {Echeverr\'ia}, \citenamefont
  {Cremades}, \citenamefont {Barrag\'an},\ and\ \citenamefont
  {Alvarez}}]{B801115J}%
  \BibitemOpen
  \bibfield  {author} {\bibinfo {author} {\bibfnamefont {B.}~\bibnamefont
  {Cordero}}, \bibinfo {author} {\bibfnamefont {V.}~\bibnamefont {G\'omez}},
  \bibinfo {author} {\bibfnamefont {A.~E.}\ \bibnamefont {Platero-Prats}},
  \bibinfo {author} {\bibfnamefont {M.}~\bibnamefont {Rev\'es}}, \bibinfo
  {author} {\bibfnamefont {J.}~\bibnamefont {Echeverr\'ia}}, \bibinfo {author}
  {\bibfnamefont {E.}~\bibnamefont {Cremades}}, \bibinfo {author}
  {\bibfnamefont {F.}~\bibnamefont {Barrag\'an}},\ and\ \bibinfo {author}
  {\bibfnamefont {S.}~\bibnamefont {Alvarez}},\ }\href
  {https://doi.org/10.1039/B801115J} {\bibfield  {journal} {\bibinfo  {journal}
  {Dalton Trans.}\ ,\ \bibinfo {pages} {2832}} (\bibinfo {year}
  {2008})}\BibitemShut {NoStop}%
\bibitem [{MgG()}]{MgGeH6}%
  \BibitemOpen
  \href@noop {} {}\bibinfo {note} {Just before submission of the manuscript, we
  noticed that the isotypic structure to the (12-1-2-1)$_V$ structure has been
  proposed by Refs.~\cite{C7CP05267G, https://doi.org/10.1002/wcms.1582} for a
  ternary hydride, $\mathrm{MgGeH}_6$.}\BibitemShut {Stop}%
\end{thebibliography}%

\end{document}